\documentclass[12pt]{article}
\usepackage{amsmath,amssymb,epsfig}
\usepackage{graphicx,floatflt,subfigure}
\usepackage{cite}

\makeatletter

\def\unit{{1\kern-.65ex {\rm l}}}
\def\1{{1\kern-.65ex {\rm l}}}

\newcount\hour \newcount\minute
\hour=\time \divide \hour by 60
\minute=\time
\def\now{%
\ifnum \hour<13
  \ifnum \hour=0 \advance \hour by 12 \number\hour:\else \number\hour:\fi%
     \ifnum \minute<10 0\fi%
     \number\minute%
\ A.M.%
\else \advance \hour by -12 \number\hour:%
  \ifnum \minute<10 0\fi%
  \number\minute%
  \ P.M.%
\fi%
}

\makeatother

\begin{document}

\baselineskip=18pt  
\numberwithin{equation}{section}  
\allowdisplaybreaks  



%
%


\thispagestyle{empty}

\vspace*{-2cm}
\begin{flushright}
 EFI-10-23\\
\end{flushright}

\vspace*{0.8cm}
\begin{center}
 {\LARGE Comments on Flipped $SU(5)$ (and F-theory)\\}
 \vspace*{1.5cm}
 E.~Kuflik$^1$, J.~Marsano$^2$\\
 \vspace*{1.0cm}
 
 $^1$ {\it Michigan Center for Theoretical Physics, University of Michigan\\
 450 Church Street, Ann Arbor, MI, 48109, USA}\\[1ex]
  $^2$ {\it Enrico Fermi Institute, University of Chicago\\
 5640 S Ellis Ave, Chicago, IL 60637, USA}
 \vspace*{0.8cm}
\end{center}
\vspace*{.5cm}

\noindent
We study the impact of nonrenormalizable operators in flipped $SU(5)$ that can generate a large $\mu$ term, R-parity violation, and rapid proton decay. While our motivation is to determine whether F-theory can naturally realize flipped $SU(5)$,   this analysis is general and leads to a characterization of symmetries capable of controlling such operators and should be independent of F-theory. We then discuss some specific implications for F-theory model building, where a significant $\mu$ problem is unavoidable.  Finally, we mention previously noted difficulties associated to engineering GUT-Higgs fields in F-theory, suggest a direct engineering of $SU(5)\times U(1)_{\chi}$ as an alternative, and present a sample construction of this type.

\newpage
\setcounter{page}{1} 



\tableofcontents


\section{Introduction}

Not long after the development of $SU(5)$ GUTs, flipped $SU(5)$ emerged as a natural alternative \cite{Barr:1981qv,Derendinger:1983aj,Antoniadis:1987dx}.  Based on gauge group $SU(5)\times U(1)_{\chi}$, flipped $SU(5)$ is not a model of unification per se, but can accommodate the near unification of couplings that is observed by experiment while overcoming difficulties of minimal $SU(5)$ models that emerged as lower bounds on the proton lifetime increased.  These successes center on the breaking of $SU(5)\times U(1)_{\chi}$ by nonzero vevs for components of "GUT-Higgs" fields that arise as a $\mathbf{10}/\mathbf{\overline{10}}$ pair. The degrees of freedom in these fields are just what is needed to lift leptoquarks and Higgs triplets from the low energy spectrum in a simple and elegant way.

In string theory, flipped $SU(5)$ models are of interest for a variety of reasons. It provides a mechanism for breaking the GUT group in 4 dimensions while solving doublet-triplet splitting without using large GUT representations. Such representations are typically unavailable in string theories. In weakly coupled Heterotic models, flipped $SU(5)$ gives one the flexibility to achieve gauge coupling unification at the string scale $\sim 10^{17}$ GeV if extra vector-like particles are added as in \cite{Jiang:2006hf}.  
In perturbative type II GUT constructions based on intersecting branes, flipped $SU(5)$ is a natural goal \cite{Chen:2005aba,Chen:2005mm,Chen:2005cf,Chen:2006ip,Cvetic:2006by} because one of the two MSSM Yukawa couplings is forced to be generated nonperturbatively there \cite{Blumenhagen:2007zk} and hence is strongly suppressed.  In perturbative type II constructions with ordinary $SU(5)$, the top Yukawa is the small one but in flipped $SU(5)$ it is the down Yukawa that is suppressed, allowing the top Yukawa to be large.

More recently,  there has been substantial interest in building flipped $SU(5)$ models in F-theory.  This includes "ultra-local" constructions \cite{Jiang:2009za,Li:2009cy} in the spirit of \cite{Beasley:2008kw}, phenomenological studies based on those constructions \cite{Li:2009fq,Li:2010mr,Li:2010dp,Li:2010rz,Li:2010ws}, and, quite recently, several "semi-local" and "global" realizations \cite{King:2010mq,Chen:2010tp,Chen:2010ts,Chung:2010bn}{\footnote{F-theory models describe physics near a stack of 8-dimensional branes in a nonperturbative background of type IIB string theory.  "Ultra-local" models are based on intuition gained by studying physics on a single coordinate patch of the brane worldvolume.  "Semi-local" models describe physics along the entire brane worldvolume and "global" models describe an embedding of the branes into a complete F-theory compactification.}}.  At first, one might think that minimal $SU(5)$ models \cite{Donagi:2008ca,Beasley:2008kw,Beasley:2008dc,Donagi:2008kj} would be more economical in this setting; there is no problem with Yukawa suppression and many problems of ordinary 4-dimensional $SU(5)$ GUTs are avoided.  Because F-theory models become effectively 8-dimensional at high scales, the GUT gauge group can be broken by turning on a nontrivial flux in the direction of $U(1)_Y$ along the internal dimensions \cite{Beasley:2008dc,Donagi:2008kj}.  This method of breaking roughly identifies the GUT scale with the compactification scale of an 8-dimensional gauge theory, hereafter referred to as the KK scale $M_{KK}${\footnote{This identification is expected to be modified slightly as a result of contributions to gauge coupling renormalization from loops of closed string fields \cite{Conlon:2009xf,Conlon:2009kt,Conlon:2009qa}.}}, and facilitates a simple removal of leptoquarks and Higgs triplets \cite{Beasley:2008kw,Donagi:2008kj}.

These successes do not come for free.  The $U(1)_Y$ flux, for instance, is known to distort gauge couplings at the KK scale \cite{Blumenhagen:2008aw,Donagi:2008kj} in a way that may be problematic.  
Further, if one tries to combine $U(1)_Y$ flux with the mechanism of \cite{Heckman:2008qa,Bouchard:2009bu,Cecotti:2009zf} for generating flavor hierarchies, light charged exotic fields necessarily appear \cite{Marsano:2009gv,Marsano:2009wr}{\footnote{As has been emphasized by J.~Heckman and C.~Vafa, this conclusion relies on the assumption of an underlying $E_8$ structure in "semi-local" F-theory GUTs as described in Appendix \ref{sec:Ftheory}.  We are not aware of any way to build semi-local or global F-theory models that avoids this so it may be that none exists.  This is far from a proof, though, and it should be stressed that finding examples that do evade this structure would be very interesting.}}.  There may be an interplay between the effects of these charged exotics, if they can be made sufficiently massive, and the distortion of unification \cite{Marsano:2009wr}.  Such a picture is significantly more complex than one might have hoped for based on the simplicity of "ultra-local" models \cite{Beasley:2008kw}, though, and therefore loses some of its appeal.  For these reasons, it is important to investigate new mechanisms for breaking the GUT group or obtaining flavor hierarchies in F-theory models.  Some promising ideas related to flavor include \cite{Font:2008id,Dudas:2009hu,King:2010mq}.  As for breaking the GUT group, flipped $SU(5)$ provides a natural alternative.

In this note, we do not focus entirely on the explicit construction of flipped $SU(5)$ models in F-theory, but rather on several phenomenological pitfalls that we encountered along the way and their implications for model building efforts.  We first study the effects of nonrenormalizable operators and different choices of symmetry whose implementation can deal with them.  This simple analysis is quite general and may be useful to see what is needed to embed flipped $SU(5)$ in a variety of string frameworks.  After that, we center the discussion on issues specific to F-theory models.
We should stress that our motivation is an alternative to GUT-breaking via hypercharge flux, to avoid disturbing gauge coupling unification. We therefore always insist that GUT-breaking and doublet-triplet splitting is accomplished via the $\mathbf{10}/\mathbf{\overline{10}}$ "GUT-Higgs" fields.  We also work entirely within the framework of "minimal" flipped $SU(5)$, wherein the only light degrees of freedom are those of the MSSM and the pair of GUT-Higgs fields{\footnote{We sometimes make reference to the addition of vector-like pairs of complete $SU(5)$ multiplets as in \cite{Jiang:2006hf}}}.  Models based on $SO(10)$ that utilize multiple fluxes to break the GUT group, as advocated for instance in \cite{King:2010mq,Chen:2010ts}, do not suffer from the problems that we will discuss but will nonetheless have to deal with certain implications of $U(1)_Y$ flux{\footnote{Since $SO(10)$ and $SU(5)\times U(1)_{\chi}$ are broken at essentially the same scale in such models, they are probably best thought of as F-theory realizations of $SO(10)$ GUTs rather than flipped $SU(5)$ "GUTs".  One might also try to engineer an $SO(10)$ model that incorporates field theoretic breaking first to $SU(5)\times U(1)_{\chi}$ and then to the MSSM.  Field theory models that do this were studied in \cite{Huang:2006nu}.}}.

\subsection{Nonrenormalizable Operators in Flipped $SU(5)$}

When flipped $SU(5)$ models are UV completed into any particular string theory framework, physics at high scales can generate nonrenormalizable operators.  Such operators can be dangerous because they arise at a scale that cannot be much larger than the roughly GUT-scale vevs of the "GUT-Higgs"  fields.  Innocent-looking operators of large dimension can therefore be transformed by the "GUT-Higgs" vevs into much more phenomenologically dangerous operators of dimension 4 and less that are not very strongly suppressed.  The role of nonrenormalizable operators has been studied before in some specific examples, such as \cite{Lopez:1990yk}, where the resulting models are quite complicated and involve many new exotic fields.  In this note, our interest is in the simplest type of flipped $SU(5)$ model, namely the one that exhibits a minimal particle content.  That is, we include only the fields of the MSSM and the "GUT-Higgs" fields needed to break the flipped $SU(5)\times U(1)_{\chi}$ gauge group.

We were not able to find an exhaustive analysis in the literature of nonrenormalizable operators in flipped $SU(5)$ models, so we undertook this exercise and characterized the types of symmetries that can lead to favorable phenomenology.  The most significant challenges are related to the $\mu$ problem, whose severity depends on one's attitude toward fine-tuning, although is should be noted that flipped $SU(5)$ was partially motivated to solve tuning problems.  Of particular importance is a dimension 7 operator that does not seem to have been discussed in the literature before.  This operator generates an enormous contribution to the $\mu$ parameter $(>10^{10}\text{ GeV})$ and can only be controlled by an $R$-symmetry{\footnote{The importance of $R$-symmetries in flipped $SU(5)$ models has been noted before \cite{Dedes:1997px,Rehman:2009yj} but we are not aware of a discussion of the operator that we study in this note.}}.  Our interest in F-theory makes this particularly troubling because "semi-local" F-theory models do not possess a suitable $R$-symmetry to deal with this. In those models, we therefore expect it to be generated and lead to a severe $\mu$ problem for which no simple solution is apparent.  This issue may be important for a wider class of UV completions of flipped $SU(5)$ in string theory as well.

In addition to this, there appears to be some tension between the $\mu$ problem and generation of neutrino masses.  Because the same Yukawa coupling that gives up-quark masses also contains the left and right handed neutrinos, it is well-known that a large Dirac neutrino mass will be generated.  Allowing the Majorana term that is needed to implement a successful type I seesaw simultaneously makes it impossible to forbid a bare $\mu$ term{\footnote{In F-theory models, the absence of a symmetry that prevents $\mu$ essentially means that $h_u$ and $h_d$ must arise as a vector-like pair of zero modes on the same matter curve.  While the presence of such vector-like pairs is rather generic when the matter curve has genus 1 or larger, there is no reason to expect that the pair remains massless since they can couple to moduli fields that can potentially acquire large vevs.}}.  One can also run into trouble with dimension 5 proton decay operators here, but there are many factors such as sparticle masses and mixings that can potentially alleviate this problem \cite{Raby:2002wc}.

Apart from the $\mu$ problem, we observe that the prevention of rapid (dimension 4-induced) proton decay requires discrete symmetries that do not have their origin as an unbroken subgroup of a continuous $U(1)$ symmetry that preserves the ordinary MSSM Lagrangian.  For us, this is unfortunate because $U(1)$'s of this type are relatively easy to engineer in $F$-theory and represent the simplest way to generate discrete symmetries in that setting.  The requisite symmetries must instead be engineered "by hand" in F-theory models as honest discrete isometries of the compactification manifold that act in the right way on the zero modes that give rise to 4-dimensional fields. 
Obtaining such symmetries is conceptually straightforward but technically challenging; the only attempt we are aware of in an F-theory context was undertaken in \cite{Hayashi:2009bt}.  

\subsection{Challenges for F-theory}

After characterizing symmetries, we then turn to some "F-theory-specific" challenges.  Here, the most serious problems are engineering the GUT-Higgs fields and explaining their vevs.  As has also been noted in the recent studies \cite{King:2010mq,Chen:2010tp,Chen:2010ts,Chung:2010bn}, it seems very difficult to engineer only the MSSM and GUT-Higgs fields in models based on $SO(10)$ without obtaining additional exotics.  The only solution seems to be realizing the GUT-Higgs as a vector-like pair which one expects to have a KK scale mass.  One must then invent a mechanism by which very massive fields manage to acquire nonzero vevs.  An alternative approach that we suggest is to build the gauge group $SU(5)\times U(1)_{\chi}$ directly.  One gives up on unification here, making the proximity of $\alpha_1$ to the other MSSM couplings at high scales seem like an accident, but at least the right spectrum of 4-dimensional fields can be realized.  To this end, engineering $SU(5)$ is straightforward but Abelian groups that do not embed into non-Abelian ones are somewhat subtle in F-theory.  Fortunately, there has been recent progress in our understanding of these $U(1)$'s  \cite{Grimm:2010ez,Marsano:2010ix} so it is possible to build compactifications for which we can reliably say that $U(1)_{\chi}$ exists as an honest gauge symmetry.  In an Appendix, we provide a simple example of a compactification of this type based on the geometries of \cite{Marsano:2009ym}{\footnote{There is no reason one has to use geometries like those of \cite{Marsano:2009ym}, which along with the compact models of \cite{Blumenhagen:2009yv,Grimm:2009yu} were constructed with the use of $U(1)_Y$ flux in mind.}}.  Several technical challenges remain, though, since we must ensure that $U(1)_{\chi}$ is not rendered anomalous by any of the fluxes that we use to induce chirality in the spectrum.  Neither this issue, nor a simple way to count the number of ($U(1)_{\chi}$-charged) $SU(5)$ singlets, are well understood at the moment.

\subsection{Outline}

The remainder of this note is organized as follows.  In section \ref{sec:flipped}, we briefly summarize the main features of flipped $SU(5)$ models.  In sections \ref{sec:dim7}-\ref{sec:Neutrino}, we study nonrenormalizable operators in flipped $SU(5)$ that are problematic for phenomenology and the symmetries that can deal with them.  Section \ref{sec:dim7} focuses on the $\mu$ problem, section \ref{sec:dim5} on $R$-parity violation, and section \ref{sec:Neutrino} on neutrino physics.  Finally, in section \ref{sec:Ftheorychallenges} we comment on the implications for F-theory model building and address other challenges unique to F-theory, such as engineering the GUT-Higgs fields.  A brief review of the 8-dimensional twisted Yang-Mills theory that essentially "defines" what we mean by semi-local F-theory models can be found in Appendix \ref{sec:Ftheory}.  We also present a sample flipped $SU(5)$ model that directly engineers $SU(5)\times U(1)_{\chi}$ in Appendix \ref{sec:compact}.

\section{Brief Review of Flipped $SU(5)$}
\label{sec:flipped}

Flipped $SU(5)$ models are distinguished by their GUT gauge group, $SU(5)\times U(1)_{\chi}$, and the identification of hypercharge as a linear combination of $U(1)_{\chi}$ and a $U(1)\subset SU(5)$.  What makes these models particularly interesting for us, though, is not the GUT gauge group itself but rather the existence of a simple, 4-dimensional mechanism for breaking $SU(5)\times U(1)_{\chi}$ down to the MSSM gauge group that lifts all non-MSSM fields that carry Standard Model charge (leptoquarks and Higgs triplets).  Only one new set of fields is needed and, quite nicely, they transform in the $\mathbf{10}$ and $\mathbf{\overline{10}}$ representations, which are easy to engineer in string theory.  Models that realize this mechanism of GUT-breaking are thus a natural alternative to consider in F-theory if one is looking for something other than internal flux to break the GUT group.  Before considering this in earnest, though, we begin in this section by reviewing how this method of GUT-breaking works and the structure of flipped $SU(5)$ models in general.

In flipped $SU(5)$, hypercharge is identified as the linear combination
\begin{equation}
q_Y= \frac{1}{5}\left(q_\chi + q_y \right),
\end{equation}
where $q_\chi$ is the $U(1)_{\chi}$ charge and $q_y$ is the $SU(5)$ hypercharge (generated by $\mbox{diag}\left(\frac{1}{3},\frac{1}{3},\frac{1}{3},-\frac{1}{2},-\frac{1}{2}  \right)$).

The MSSM matter fields and Higgs doublets transform under the $SU(5)\times U(1)_{\chi}$ as the representations
\[ F_i \equiv 10_{-1} = (Q_i, d^c_i, \nu^c_i) \;\;\;\;  \bar{f}_i \equiv \overline{5}_{3} = (u^c_i, L_i)  \;\;\;\;  \ell_i \equiv 1_{-5} = (e^c_i) \]
\begin{equation}
h \equiv 5_{2} = (D_h, h_d) \;\;\;\; \bar{h} \equiv \overline{5}_{-2} = (\bar{D}_h, h_u) \label{MSSMHiggs}
\end{equation}
where $i$ is a family index. Notice the ``flipped'' assignments of  $d^c-u^c$, $e^c-\nu^c$ and $h_u-h_d$ in comparison to their typical assignment in the Georgi--Glashow $SU(5)$  model. The matter and Higgs fields participate in the typical Yukawa couplings
\[ W \supset y_d^{ij}   F_i   F_j   h + y_u^{ij}   F_i   \bar{f}_j   \bar{h} +  y_e^{ij}   \bar{f}_i    \ell_j   h,\]
Note in ``flipped'' models the charged lepton and down-type quark masses need not unify, but Dirac neutrino masses and up-type quark masses do unify.

To break $SU(5)\times U(1)_{\chi}$, one introduces two new GUT-Higgs fields
\begin{equation}
H \equiv 10_{-1} = (Q_H, D^c_H, \nu^c_H) \;\;\;\; \bar{H} \equiv \overline{10}_{1} = (\bar{Q}_H, \bar{D}^c_H, \bar{\nu}^c_H)
\end{equation}
whose vacuum expectation values are aligned in the SM neutral directions $\langle \nu^c_H \rangle = \langle \bar{\nu}^c_H \rangle \sim M_{GUT}.$  Leptoquarks are removed via the super-Higgs mechanism.  To deal with Higgs triplets, one includes the superpotential couplings
\begin{equation}
W_{\mbox{Flipped}} = \lambda_H   H   H   h +  \bar{\lambda}_{H}   \bar{H}   \bar{H}   \bar{h} \label{flippedW}
\end{equation}
which give masses to the Higgs color triplets
\begin{equation}
\lambda_H   \langle  \nu^c_H \rangle   D^c_H   D_h + \bar{\lambda}_H   \langle  \bar{\nu}^c_H   \rangle   \bar{D}^c_H   \bar{D}_h \label{TripMass}
\end{equation}
via mixing with the triplet components of the $H,\bar{H}$ multiplets. Since (\ref{flippedW}) leaves the Higgs doublets massless, flipped $SU(5)$ solves the doublet-triplet splitting problem.

The scale at which $SU(5)\times U(1)_{\chi}$ breaks to the MSSM is set by the $\nu_H^c,\bar{\nu}_{\bar{H}}^c$ expectation values and is typically referred to as $M_{32}$, since only the $SU(3)_c$ and $SU(2)_L$ couplings need unify there.  Unless the Higgs triplets are anomalously light due to small values of $\lambda_H$ or $\bar{\lambda}_H$, $M_{32}$ sits near the typical GUT scale $M_{\rm GUT}\sim 2\times 10^{16}$ GeV.

As outlined in the introduction we will think of flipped $SU(5)$ as originating from an underlying $SO(10)$ theory.  The scale at which $SU(5)$ and $U(1)_{\chi}$ unify into $SO(10)$ is typically referred to as the "super-unification" scale and denoted by $M_{su}$.  In F-theory realizations that break $SO(10)$ to $SU(5)\times U(1)_{\chi}$ with an internal flux, $M_{su}$ also denotes the Kaluza-Klein scale above which the physics becomes effectively 8-dimensional.  For this reason, it provides us with a natural cutoff scale for 4-dimensional physics above which we expect towers of new states to generate nonrenormalizable couplings.  We will use $\Lambda$ to denote the cutoff scale of our 4-dimensional theory in the remainder of this note in order to be as general as possible, keeping in mind $\Lambda \simeq M_{su}$ in a large class of theories.

\subsection{The Need for Symmetries}

Nonrenormalizable couplings containing the GUT-Higgs fields can be a potentially serious problem in flipped $SU(5)$ models.  In the presence of the large, nonzero expectation values of $\nu_H^c$ and $\bar{\nu}_H^c$,  these can give rise to renormalizable couplings involving only MSSM fields.  Such operators will typically be suppressed by powers of $M_{32}/\Lambda$ where $\Lambda$ is the cutoff scale at which the operator is generated that we usually take to be $M_{su}$.  In the rest of this note we will denote the suppression factor by $\delta$
\begin{equation}\delta\equiv \frac{\langle \nu_H^c \rangle}{ \Lambda} \sim \frac{M_{32}}{g \Lambda}\end{equation}
where we used that the GUT-Higgs vev is related to the unfication scale by, $\langle \nu_H^c \rangle \sim M_{32}/g $ with $g$ the $SU(5)$ coupling constant at $M_{32}$.

In general, $\delta$ cannot be too small.  To get a conservative estimate for it, we can first effectively replace $\Lambda$ by $M_{\text{Planck}}${\footnote{One can in principle raise $M_{su}$ up to $M_{\text{Planck}}$ by introducing new vector like pairs at the TeV scale or above as in  \cite{Jiang:2006hf}.}} .  As for $M_{32}$, this is not quite the standard MSSM unification scale $M_{\rm GUT}\sim 2\times 10^{16}$ GeV because the triplets, being somewhat lighter than $M_{32}$, contribute to the running.  However, the unification scale $M_{32}$ can be calculated from the 1-loop RGE equations for the flipped $SU(5)$ matter content,  thus relating the triplet masses  $m_{T,H}$ and $m_{T,\bar{H}}$, the scales $M_{32}$ and $M_{\rm GUT}$  by
\begin{equation}M_{32}^2\simeq \frac{M_{\rm GUT}^4}{m_{T,H}m_{T,\bar{H}}}\label{M32trip}\end{equation}
This means that the $SU(3)-SU(2)$ unification scale is actually increased relative to $M_{\rm GUT}$ by the triplets{\footnote{If the triplets become heavier than $M_{\rm GUT}$ then \eqref{M32trip} indicates $M_{32}<M_{\rm GUT}$.  In that case, though, the triplets would be heavier than $M_{32}$ so would not contribute to the running.  In fact, we have a bigger problem if $m_{T,a}$ is much larger than $M_{32}$ because this would mean that the theory at $M_{32}$ is becoming strongly coupled.  We will always assume perturbativity, and hence $m_{T,a}<M_{32}$, leading to $M_{32}>M_{\rm GUT}$.}} allowing us to replace $M_{32}$ by $M_{\rm GUT}$ to get a conservative estimate for $\delta$.  Putting it all together, we find that
\begin{equation}\delta\gtrsim \frac{M_{\rm GUT}}{M_{\text{Planck}}}\sim 10^{-2}.\end{equation}
For many operators, this suppression will be entirely insufficient.
 
To control the effects of problematic  nonrenormalizable operators, we must therefore introduce new symmetries.  A drawback of flipped $SU(5)$ models is that one of the most useful symmetries for forbidding unwanted operators in the MSSM, $U(1)_{\chi}$, is strongly broken by the vevs of the GUT-Higgs fields.  One reason that $U(1)_{\chi}$ is often so useful is that, as is well-known, it contains matter parity as a $\mathbb{Z}_2$ subgroup.  Unfortunately, the GUT-Higgs fields carry odd $U(1)_{\chi}$ charge so not even this nice $\mathbb{Z}_2$ subgroup remains after GUT-breaking.

The situation is in fact worse because both $F$ and $H$ carry identical charges under any continuous symmetry that preserves the full Yukawa and flipped superpotential
\begin{equation}W_{\text{Yukawa}+\text{Flipped}} \sim FFh +  F\bar{f}\bar{h} + \bar{f}\ell h + HHh + \bar{H}\bar{H}\bar{h}\label{MSSMflipped}\end{equation}
Any attempt to realize matter parity as a subgroup of a continuous symmetry is bound to fail; $H$ will always have the same parity as $F$, that is odd parity, and break it spontaneously.  This is important for building string models because it means that matter parity must always be engineered on its own as an honest discrete symmetry.

Since $U(1)_{\chi}$ and its famous $\mathbb{Z}_2$ subgroup are unavailable, we must look to other options.  In this note, the symmetries that we shall consider are of three types: discrete $\mathbb{Z}_n$ symmetries, continuous $U(1)$ symmetries, and $U(1)_R$ symmetries.  The charges of all fields under the most general $\mathbb{Z}_n$, $U(1)$, and $U(1)_R$ symmetries that are consistent with \eqref{MSSMflipped} are listed in Table \ref{chargetable},
\begin{table}[h!]
 \[\begin{array}{c|ccc}
\text{Field} & \mathbb{Z}_n\text{ charge} & U(1)\text{ charge} & U(1)_R\text{ charge} \\ \hline
H & r & p & p_R \\
\bar{H} & s & q & q_R \\
h & -2r\text{ mod }n & -2p & -2p_R+2 \\
\bar{h} & -2s\text{ mod }n & -2q & -2q_R+2 \\
F & r+\epsilon\frac{n}{2}\text{ mod }n & p & p_R \\
\bar{f} & 2s-r+\epsilon\frac{n}{2}\text{ mod }n & 2q-p & 2q_R-p_R \\
\ell & 3r-2s+\epsilon\frac{n}{2}\text{ mod }n & 3p-2q & 3p_R-2q_R
\end{array}\]
\caption{\label{chargetable} All symmetries consistent with the full Yukawa + flipped superpotential (\ref{MSSMflipped}), where $r$ and $s$ are taken to lie between 0 and $n-1$ and $\epsilon$ can take the value 0 or 1 if $n$ is even but must be zero if $n$ is odd. }
 \end{table}
where $r$ and $s$ are taken to lie between 0 and $n-1$.  The parameter $\epsilon$ can take the value 0 or 1 if $n$ is even but, obviously, must be zero if $n$ is odd.  Two common symmetries that appear in the literature are matter parity and a $\mathbb{Z}_2$ that goes by the name of $H$-parity.  In the language of Table \ref{chargetable}, these correspond to
\begin{equation}\mathbb{Z}_2^{(\text{Matter parity})}\leftrightarrow n=2,\,\, \epsilon=1,\,\,r=s=0\label{mattpar}\end{equation}
and
\begin{equation}\mathbb{Z}_2^{(H\text{-parity})}\leftrightarrow n=2,\,\,\epsilon=1,\,\,r=1,\,\,s=0\label{hpar}\end{equation}

In the next few sections, we will discuss ways to use symmetries of these types to address the $\mu$ problem, $R$-parity violation, and dimension 5 proton decay while simultaneously generating small neutrino masses{\footnote{We perform an operator analysis rather than studying which symmetries can be left unbroken by vevs of $H$ and $\overline{H}$ because, in most cases, we do not need to forbid operators per se; we only need to suppress them.  Often, a $\mathbb{Z}_n$ symmetry with a sufficiently large value of $n$ will be sufficient even though what remains of it after GUT-breaking does not forbid anything}}.  The issue of controlling higher dimension operators in flipped $SU(5)$ models is of course not new but we are unaware of any previous work regarding some of the operators that we study.  This is particularly true for the most troublesome operator, which has dimension 7 and will be studied in section \ref{sec:dim7}.

\section{The $\mu$ Problem}
\label{sec:dim7}

\subsection{Generalities}

We begin by studying the generation of the supersymmetric Higgs mass, $\mu$
\begin{equation}W_{\mu}\sim \mu h\bar{h}\end{equation}
In addition to this bare $\mu$ term, there is an entire tower of operators that can generate a nonzero $\mu$ after GUT-breaking
\begin{equation}\frac{1}{\Lambda^{2m-1}}\left(H\bar{H}\right)^m h\bar{h}\supset \left(\frac{\langle \nu_H^c \bar{\nu}_{H}^c \rangle}{\Lambda^2}\right)^m h\bar{h}\rightarrow g^{-1}\delta^{2m-1}M_{32}h\bar{h}\label{muoperators}\end{equation}
The charges of these operators under the symmetries in Table \ref{chargetable} are
\begin{equation}
\begin{array}{c|ccc}
\text{Operator} & \mathbb{Z}_n\text{ charge} & U(1)\text{ charge} & U(1)_R\text{ charge} \\ \hline
\Lambda^{-(2m-1)}(H\bar{H})^m h\bar{h} & (m-2)(r+s)\text{ mod }n & (m-2)(p+q) & (m-2)(p_R+q_R)+4
\end{array}\end{equation}

Any continuous $U(1)$ symmetry that forbids a bare $\mu$ term has $p+q\ne 0$ and succeeds in forbidding all operators in the tower with $m\ne 2$.  If we are interested in generating $\mu$ but ensuring that it is suppressed, we can instead try to use a $\mathbb{Z}_n$ symmetry with sufficiently large $n$ since, for suitable values of $r$ and $s$, the first solution other than $m=2$ will sit at $m=n+2$.  With $M_{32}\sim 10^{16}$ GeV and $\delta\sim 10^{-2}$, the operator with $m=4$ will generate a $\mu$ of the right size $\sim 10^2$ GeV.  A suitable $\mathbb{Z}_2$ symmetry is sufficient to forbid $m=1$ and $m=3$.

Unfortunately, the non-$R$ symmetries always have a problem with the dimension 7 operator at $m=2$ which, to our knowledge, has not been discussed previously in the literature
\begin{equation}{\cal{O}}_7 = \frac{1}{\Lambda^3}(H\bar{H})^2(h\bar{h})\label{O7def}\end{equation}
It is easy to see why $\mathbb{Z}_n$ and $U(1)$ have trouble forbidding this.  The charge of ${\cal{O}}_7$ under any non-$R$ symmetry is the sum of charges of two terms, $hHH$ and $\bar{h} \bar{H} \bar{H}$, that are needed to lift the Higgs triplets.  The only way to control it, then, is with an $R$-symmetry.  More specifically, any $U(1)_R$ with $p_R+q_R$ neither 1 nor 2 is sufficient to eliminate the entire tower, including ${\cal{O}}_7$.

As discussed in Appendix \ref{sec:Ftheory}, though, the underlying 8-dimensional gauge theory of F-theory models does not provide a suitable $R$-symmetry so one always expects the operator ${\cal{O}}_7$ \eqref{O7def} to be generated.  For this reason, we will spend a little more time studying it.  The problem with ${\cal{O}}_7$ is that the $\mu$ term it induces is enormous
\begin{equation}\mu_{\text{induced}}\gtrsim g^{-1}\delta^3M_{32}\sim 10^{10}\text{ GeV}\label{muest}\end{equation}
This introduces an enormous fine-tuning problem for electroweak symmetry breaking, which defeats the purpose of building a Flipped $SU(5)$ to solve the tuning related to doublet-triplet splitting.

Recall that this estimate, which is based on taking $\Lambda\sim M_{\text{Planck}}$, is particularly conservative if we insist on realizing flipped $SU(5)$ in a semi-local F-theory model; reliability of the entire semi-local approach depends on having control over the underlying 8-dimensional gauge theory which, in turn, requires $\Lambda$ to be at least an order of magnitude or two smaller than $M_{\text{Planck}}$.

\subsection{Suppressing ${\cal{O}}_7$ with an approximate non-R symmetry}

We cannot expressly forbid ${\cal{O}}_7$ \eqref{O7def} with a global non-R symmetry without losing the ''flipped superpotential'' \eqref{flippedW} but we can imagine trying to suppress it with an approximate symmetry that is spontaneously broken.  Since (\ref{flippedW}) must be generated if flipped $SU(5)$ is to elicit doublet-triplet splitting, the couplings in (\ref{flippedW}) are replaced by

\begin{equation}
W \supset \dfrac{S}{\Lambda} H H h + \dfrac{\bar{S}}{\Lambda} \bar{H}   \bar{H}   \bar{h}. \label{NewFlippedW}
\end{equation}
for some fields $S$  and $\bar{S}$. Passing to expectation values, we define the dimensionless quantities $\lambda_H$ and $\bar{\lambda}_H$ as
\begin{equation}
 \lambda_H \sim \frac{\langle S \rangle}{\Lambda} \;\;\;\;  \bar{\lambda}_{H} \sim \frac{\langle \bar{S} \rangle}{\Lambda}.
\end{equation}
Since the product of couplings in (\ref{NewFlippedW}) is an invariant, at the very least the operator ${\cal{O}}_7$ \eqref{O7def} will be generated with suppression of $ \lambda_H \bar{\lambda}_{H} $,
\begin{equation}
{\cal{O}}_7' = \lambda_H \bar{\lambda}_{H}  \frac{(H \bar{H})^2}{\Lambda^3} h \bar{h} \quad\rightarrow \quad\lambda_H\bar{\lambda}_H\delta^3g^{-1}M_{32}\,\int\,d^2\theta\,h\bar{h} \label{NewMu}.
\end{equation}

In this case, it naively seems that $\mu$ can be less than our previous estimate \eqref{muest} if  $ \lambda_H ,\bar{\lambda}_{H} \ll 1$. However, as $\lambda_H ,\bar{\lambda}_{H}$ are lowered, the Higgs triplet masses, given in (\ref{TripMass}), will also be lowered. From \eqref{M32trip}, then, we see that the scale $M_{32}$ becomes larger as we do this, and may even become super-Planckian.  Taken together, it is in fact easy to see that all dependence of \eqref{NewMu} on $\lambda_H$ and $\bar{\lambda}_H$ cancels completely.  This is because the triplet masses are related to $\lambda_H$ and $\bar{\lambda}_H$ by
\begin{equation}m_{T,H} \simeq \frac{\lambda_H}{g} M_{32} \qquad m_{T,\bar{H}} \simeq \frac{\bar{\lambda}_H}{g} M_{32} \end{equation}
Using this, \eqref{M32trip} becomes
\begin{equation}M_{32}^4 \sim M_{\rm GUT}^4 \times \frac{g^2}{\lambda_H\bar{\lambda}_H}\end{equation}
which leads to an induced $\mu$ term
\begin{equation}\mu_{\text{induced}}\sim \frac{\lambda_H\bar{\lambda}_H}{g^4}\frac{M_{32}^4}{\Lambda^3}\sim g^{-2}\left(\frac{M_{\rm GUT}^4}{\Lambda^3}\right)> 10^{10}\,\text{GeV}\end{equation}
Introducing an approximate symmetry is, perhaps counterintuitively, not effective at lowering $\mu_{\text{induced}}$. 

\subsection{Summary}

The only way to avoid generating any contribution to the $\mu$ term after GUT-breaking is with a $U(1)_R$ symmetry that has $p_R+q_R\ne 2$.  In the absence of such a symmetry, one expects ${\cal{O}}_7$ \eqref{O7def} to appear and lead to a $\mu$ term that is far too large.  Provided a solution to the ${\cal{O}}_7$ problem can be found, a continuous $U(1)$ symmetry with $p+q\ne 0$ can get rid of the remaining operators in \eqref{muoperators} while a $\mathbb{Z}_2$ can allow only those that give rise to $\mu\sim 10^2$ GeV or smaller.  One idea for solving the ${\cal{O}}_7$ problem revolves around forbidding the terms \eqref{flippedW} that generate masses for Higgs triplets with a continuous symmetry and breaking it through the vevs of suitable singlet fields.  This solution does not appear to work, however, so one needs something more intricate.

\section{$R$-Parity Violating Operators}
\label{sec:dim5}

Putting the $\mu$ problem aside for now, we next turn our attention to the generation of renormalizable MSSM superpotential couplings that violate $R$-parity.  These couplings take the form
\begin{equation}W_{\not R} \sim \lambda LLe^c + \lambda' QLd^c + \lambda^{\prime\prime}u^c d^c d^c + \kappa Lh_u\label{Rparviol}\end{equation}
The coupling $\kappa$ can be rotated away by a field redefinition but only at the cost of inducing new contributions to the lepton violating trilinear couplings.

It is well-known that $U(1)_{\chi}$ contains a $\mathbb{Z}_2$ that acts like matter parity on MSSM fields, which means that none of the operators in \eqref{Rparviol} can arise on their own in a flipped $SU(5)$ model, which is based on gauge group $SU(5)\times U(1)_{\chi}$.  However, the GUT-Higgses, $H$ and $\bar{H}$, are parity-odd and will spontaneously break this $\mathbb{Z}_2$.  Operators appearing in \eqref{Rparviol} can therefore appear in combination with suitable powers of $H$ and $\bar{H}$
\begin{equation}\begin{split}
\frac{1}{\Lambda^{2m}}(H\bar{H})^mH\bar{f}\bar{h} &\supset \left(\frac{\nu_{H}^c \bar{\nu}_{H}^c}{\Lambda^2}\right)^m \nu_{H}^cLh_u\\
\frac{1}{\Lambda^{2m+1}}(H\bar{H})^mHFF\bar{f} &\supset \left(\frac{\nu_{H}^c \bar{\nu}_{H}^c}{\Lambda^2}\right)^m\frac{\nu_{H}^c}{\Lambda}\left\{ \begin{array}{l}QDL\\ UDD\end{array}\right. \\
\frac{1}{\Lambda^{2m+1}}(H\bar{H})^mH\bar{f}\bar{f}\ell &\supset \frac{\nu_{H}^c}{\Lambda}LLE
\end{split}\label{Rviol}\end{equation}

In general, $R$-parity violating operators must be significantly suppressed, if not outright forbidden.  As we have seen, the suppression factor $\delta=(\langle H_{\nu}\rangle /\Lambda)$ is not very small, taking values $\delta\gtrsim 10^{-2}$.  This means that only operators with fairly high powers of $m$ are safe.  Additional symmetries are needed to forbid or suppress the rest.  

The charges of the $R$-parity violating operators \eqref{Rviol} under the symmetries of Table \ref{chargetable} are
\begin{equation}\begin{array}{c|ccc}
\text{Operator} & \mathbb{Z}_n\text{ charge} & U(1)\text{ charge} & U(1)_R\text{ charge} \\ \hline
\Lambda^{-2m}(H\bar{H})^m H\bar{f}\bar{h} & m(r+s)+\epsilon\frac{n}{2}\text{ mod }n & 0 & 2 \\
\Lambda^{-(2m+1)}(H\bar{H})^m HFF\bar{f} & (m+2)(r+s)+\epsilon\frac{n}{2}\text{ mod }n & (m+2)(q+p) & (m+2)(q_R+p_R) \\
\Lambda^{-(2m+1)}(H\bar{H})^m H\bar{f}\bar{f}\ell & (m+2)(r+s)+\epsilon\frac{n}{2}\text{ mod }n & (m+2)(q+p) & (m+2)(q_R+p_R)
\end{array}\end{equation}

Notice that continuous symmetries alone are not sufficient to prevent a bilinear coupling $\kappa LH_u$ with $\kappa\sim M_{32}\gtrsim 10^{16}\,\text{ GeV}$.  For this, we need at least one discrete symmetry.

\subsection{A Single $\mathbb{Z}_n$ Symmetry}

We begin then by discussing the simplest possibility, namely controlling $R$-parity violating couplings with only a single $\mathbb{Z}_n$ symmetry.  Which operators are generated depends on the set of solutions for $m$ to the equation
\begin{equation}m(r+s)+\epsilon\frac{n}{2}=0\text{ mod }n\label{meqn}\end{equation}
The simplest way to limit the number of solutions to \eqref{meqn} is to take $r+s=0$ and $\epsilon=1$.  In this case, there are no solutions and all operators in \eqref{Rviol} are expressly forbidden.  If we set $n=2$ and $\epsilon=1$, for instance, we find two $\mathbb{Z}_2$ symmetries of this type.  One of these is ordinary matter parity, $\mathbb{Z}_2^{(\text{Matter Parity})}$ \eqref{mattpar}.  There is another $\mathbb{Z}_2$ that does the job, though, under which all MSSM fields are even while the GUT-Higgs fields are odd.  It is easy to see that neither of these $\mathbb{Z}_2$'s can be embedded into a $U(1)$ symmetry that preserves the MSSM and flipped superpotentials {\footnote{This follows imediately because the operator $H\bar{f}\bar{h}$ is neutral under any such $U(1)$ but carries odd parity under each $\mathbb{Z}_2$.}}.  Note also that the commonly used $H$-parity, $\mathbb{Z}_2^{(H\text{-Parity})}$ \eqref{hpar}, allows numerous solutions starting at $m=1$ so it unable to prevent problematic $R$-parity violation.

We now investigate the possibility that $r+s\ne 0$ mod $n$.  In this case, $m=n$ is always a solution for $\epsilon=0$ while, for $\epsilon=1$, we will always get a solution at one of $m=n/2$ or $m=n$.  This is not a problem, though, provided $n$ is sufficiently large that only operators from \eqref{Rviol} with sufficient suppression are generated.

If we are given a solution $m_0$ to \eqref{meqn} then we generate the couplings $\lambda$, $\lambda'$, $\lambda^{\prime\prime}$, and $\kappa$ in \eqref{Rparviol} with the suppressions
\begin{equation}\lambda,\lambda',\lambda^{\prime\prime}\sim \delta^{2m_0-3}\qquad\kappa\sim\delta^{2m_0}M_{32}\end{equation}
where as usual $\delta = M_{32}/g\Lambda\gtrsim 10^{-2}$.  Bounds from proton decay can be model dependent but the analysis of low energy SUSY in \cite{FileviezPerez:2004th} suggests the order of magnitude constraint
\begin{equation}\lambda^{\prime}\lambda^{\prime\prime}\lesssim 10^{-24}\quad \implies m_0\ge 5\end{equation}
From $\kappa$, however, we obtain an induced contribution to $\lambda'$ that scales like $\kappa/\mu$, where $\mu$ is the supersymmetric Higgs mass.  This means that we also need
\begin{equation}\frac{\kappa}{\mu}\lambda^{\prime\prime}\gtrsim 10^{-24}\quad \implies m_0 \ge \frac{15}{4}+\frac{1}{8}\ln\left(\frac{M_{32}}{\mu}\right)\end{equation}
For $\mu\sim 100\text{ GeV}$ and $M_{32}\sim 10^{16}\text{ GeV}$, this leads to the tighter constraint
\begin{equation}m_0\ge 6\end{equation}
One would therefore need at least a $\mathbb{Z}_6$ symmetry to do the job.

\subsection{Adding a $U(1)$ Symmetry}

If we add a $U(1)$ symmetry to our $\mathbb{Z}_n$ then one can completely evade the proton decay constraints.  This is because a $U(1)$ symmetry with $q+p\ne 0$ forbids $\lambda^{\prime\prime}UDD$, which is the only source of baryon number violation in \eqref{Rparviol}.  We will still need a $\mathbb{Z}_n$ symmetry to suppress $\kappa$, though.  Operators with $m=3,4,5$ generate $\kappa$'s of order $10^4$, $1$, and $10^{-4}$ GeV, respectively. From electroweak symmetry breaking considerarions $\kappa\sim 10^{4}$ GeV is much too large, but $1$ GeV and $10^{-4}$ GeV may be ok.  In each case, $\kappa/\mu\ll 1$ for $\mu\sim 100$ GeV so it seems sensible to rotate $\kappa$ away, effectively replacing it with $\lambda'\sim \kappa/\mu$.  Bounds on $\lambda$, $\lambda'$, and $\lambda^{\prime\prime}$ individually rather than their products can be found in \cite{Allanach:1999ic}.  These bounds are model dependent, but it seems that $\lambda'\lesssim 10^{-3}$ is reasonable, leading to
\begin{equation}\kappa\lesssim 10^{-3}\mu\sim 10^{-1}\text{ GeV}\end{equation}
The $m=4$ operator seems troublesome but the operators with $m\ge 5$ should be ok.  Achieving this requires a $\mathbb{Z}_n$ symmetry with $n\ge 5$.  This is only a marginal improvement on the condition $n\ge 6$ that was necessary in the absence of a $U(1)$ symmetry.  Introducing a $U(1)$ therefore doesn't seem to buy us very much.

The story is similar for $U(1)_R$ symmetry.  We can forbid the trilinear couplings by taking $q_R+p_R>1$ but a $\mathbb{Z}_n$ symmetry with $n$ at least 5 is still needed. 

\subsection{Summary}

Continuous $U(1)$ and $U(1)_R$ symmetries are insufficient to prevent severe $R$-parity violation in conflict with the measured proton lifetime.  Discrete symmetries are necessary, with conventional $\mathbb{Z}_2$ matter parity one of the two most simple options.  For discrete symmetries that only suppress quadratic and trilinear $R$-parity violation at low energies without expressly forbidding it, the order of the group can be slightly reduced if it is combined with a continuous $U(1)$ or $U(1)_R$ symmetry.  The net effect of the continuous symmetries does not help us very much, though, so for our purposes we will treat $R$-parity violation as a problem that must be addressed by discrete symmetries.

\section{Dimension 5 Operators: Neutrino Masses and Proton Decay}
\label{sec:Neutrino}
In Flipped $SU(5)$, Dirac neutrino masses
\begin{equation}
 h_u   L   \nu^c
\end{equation}
arise from the Yukawa couplings
\begin{equation}
\bar{h}   \bar{f}   F \supset  h_u   L   \nu^c + h_u Q u^c . \label{upneutryukawa}
\end{equation}
Since (\ref{upneutryukawa}) also supplies the up-quark masses, this limits
the suppression that the
Dirac neutrino masses can have, and thus Flipped $SU(5)$ requires a
seesaw mechanism to generate small neutrino masses. Correspondingly,
the right-handed neutrino Majorana mass
\begin{equation}
 \nu^c   \nu^c \in F   F\label{rhmaj}
\end{equation}
must be present. But since $F F$ is not an $SU(5)$ invariant, the
above term originates from the non-renormalizable operator
\begin{equation}
W_{\text{Neutrino}}  = \frac{1}{\Lambda} \bar{H}   F   \bar{H}   F . 
\label{neutrinocoup}
\end{equation}
This is just the Type-I seesaw mechanism, and in terms of the scale $\Lambda$ the light neutrino masses are
\begin{equation}
 m_{\nu} = \frac{(y_u   \langle h_u \rangle)^2}{\langle \bar{\nu}^c_H \rangle^2 / \Lambda} \lesssim \frac{m_{top}^2}{\delta M_{32}}.
\label{neutrinomass}
\end{equation}

The MINOS experiment \cite{Adamson:2008zt} on neutrino oscillations is consistent with the mass splitting of two neutrino mass eigenstates,  $|\Delta m^2| = (2.43 \pm .13) \times 10^{-3} \mbox{ eV}^2$. Requiring that the heaviest neutrino be of order the mass splitting in order to minimize the tuning in the neutrino mass matrix, (\ref{neutrinomass}) gives the correct neutrino mass for $\delta \sim 10^{-2}$. Since it has already been argued that $ \delta \gtrsim 10^{-2}$, the operator (\ref{neutrinocoup}) should not be further suppressed. Therefore, a {\it necessary} condition to generate small neutrino
masses without introducing tuning into the neutrino sector is that the theory be able to generate \eqref{rhmaj} with only the suppression induced from \eqref{neutrinocoup}.

Requiring that the Majorana mass (\ref{neutrinocoup}) be invariant in addition to the flipped superpotential and Yukawa couplings in (\ref{MSSMflipped}), imposes the additional constraints 
\begin{equation}\begin{array}{ccl}
2r + 2s &=& 0 \mod n \\
2q+2p &=& 0\\
2q_R+2p_R &=& 2
\end{array}\end{equation}
on the charges in Table \ref{chargetable}. The new charges consistent with all superpotential couplings are give in Table \ref{chargetablewithneutrino}.
\begin{table}[h!]
\[\begin{array}{c|ccc}
\text{Field} & \mathbb{Z}_n\text{ charge} & U(1)\text{ charge} & U(1)_R\text{ charge} \\ \hline
H & r & p & p_R \\
\bar{H} & -r & -p & -p_R+1 \\
h & -2r\text{ mod }n & -2p & -2p_R+2 \\
\bar{h} & 2r\text{ mod }n & 2p & 2p_R  \\
F & r+\epsilon\frac{n}{2}\text{ mod }n & p & p_R \\
\bar{f} & -3r+\epsilon\frac{n}{2}\text{ mod }n & -3p & -3p_R +2 \\
\ell & 5r+\epsilon\frac{n}{2}\text{ mod }n & 5p & 5p_R -2.
\end{array}\]
 \caption{\label{chargetablewithneutrino} Same as Table \ref{chargetable}, but with the additional constraint that the Majorana neutrino mass  (\ref{neutrinocoup}) be invariant.} \end{table}

 The global $U(1)$ symmetry is exactly $U(1)_{\chi}$ up-to a scaling, so the remaining two symmetries classify all possible (Abelian, non-family) symmetries consistent with Flipped $SU(5)$ {\footnote{A discrete $R$-symmetry is also possible, but will not change the discussion below.}}.  The $\mathbb{Z}_n$ symmetry is sufficient to forbid $R$-parity violating operators provided $\epsilon=1$ while the $U(1)_R$ is enough to avoid the generation of $\mu$ from nonrenormalizable operators involving $H$ and $\overline{H}$.

\subsection{$\mu$ Problem and Dimension 5 Proton Decay}

Unfortunately, the symmetries from Table \ref{chargetablewithneutrino} cannot forbid the bare $\mu$-term
\begin{equation}
 \bar{h}h \supset h_u h_d  
\label{baremuterm}
\end{equation}
or dimension 5 proton decay operators
\begin{equation}
F F F \bar{f} + F \bar{f} \bar{f} \ell \supset QQQL + d^c u^c u^c e^c + u^c d^c d^c \nu^c
\label{dim5protondecay}
\end{equation}

Consequently,  one of the couplings in  (\ref{MSSMflipped}) and  (\ref{neutrinocoup}) needs to be forbidden if the $\mu$-term (\ref{baremuterm}) and dimension-5 proton decay operators (\ref{dim5protondecay}) are to be suppressed.  We focus here only on the $\mu$ problem as several factors can affect proton decay that could in principle be tuned \cite{Raby:2002wc}.
One can consider symmetries that forbid $\mu$ but are spontaneously broken, thereby allowing \eqref{MSSMflipped} and \eqref{neutrinocoup} to arise.

A set of superpotential operators can be distinguished
\begin{equation}
W \supset \bar{H}   \bar{H}   \bar{h} + F   F   h + \frac{1}{\Lambda}  \bar{H}   F   \bar{H}   F \label{absentcoupling}
\end{equation}
that if invariant will lead to an invariant $\mu$ term; in other words, when one is forbidden then the $\mu$ term can be forbidden. If one of the trilinear terms in (\ref{absentcoupling}) is absent, then it will necessarily have the same quantum numbers as the $\mu$ term. Then when the trilinears are generated via spontaneous symmetry breaking, so will the $\mu$-term,  and so the two will undergo similar suppression. If, on the other hand, one forbids $ \bar{H}   F   \bar{H}   F $, then this operator will have opposite charge than $h \bar{h}$. Then if the neutrino Majorana mass is generated dynamically, via an $SU(5)$  singlet $S$
\begin{equation}
\frac{S}{\Lambda^2}   \bar{H}   F   \bar{H}   F 
\label{neutrinocoupS}
\end{equation}
then the $\mu$-term is not generated by the vev of $S$. The value  $\langle S \rangle$, which feeds into (\ref{neutrinomass}), needs to be close to the scale $\Lambda$, to give adequately small neutrino masses. Generating additional GUT-sized vevs will necessarily create tension when building a successful flipped $SU(5)$ model that solves the neutrino mass problem.

One can also consider models that generate effectively $\bar{H}   F   \bar{H}   F $ when heavy fields are integrated out. The typical seesaw mechanism in Flipped $SU(5)$ \cite{Antoniadis:1987dx} comes from the renormalizable superpotential couplings
\begin{equation}
 y_u^{ij}  F_i  \bar{f}_j ~  \bar{h} + \lambda^{\nu}_{ij} \bar{H} F^i S^j + M^{S}_{ij} S^i S^j 
\label{flippedneutrino}
\end{equation}
where there are now three  $SU(5)\times U(1)_{\chi}$ singlets.  This generates the $9\times9$ neutrino mass matrix
\[\left(
\begin{array}{ccc}
 0 & y^u  \langle \overline{h} \rangle  & 0 \\
 y^u \langle \overline{h} \rangle  & 0 & \lambda^S \langle \bar{\nu}^c_H \rangle  \\
 0 & \lambda^S \langle \bar{\nu}^c_H \rangle  & M^S
\end{array}
\right)\]
 in the $( L , \nu^c, S)$ basis. Assuming $M^S \sim \langle \bar{\nu}^c_H \rangle \sim M_{32}$ this generates light neutrino masses
\begin{equation}
m_{\nu} \lesssim \frac{\langle \overline{h} \rangle^2}{M_{32}}
\end{equation}
with the same results as the Type-I seesaw mechanism described above, but without the additional factors of $\delta$.  Unfortunately, $M^{S}$ has the same quantum numbers as $\mu$, so both couplings should be of similar size, and the $\mu$ problem remains. One would need to add additional symmetries and $SU(5)$ singlet fields to make this model work. 

\subsection{Summary}

Engineering neutrino masses that do not involve fine-tuning restricts the available symmetries, making it impossible to forbid either a bare $\mu$ term or operators that lead to dimension 5 proton decay.  As it is a renormalizable coupling, the presence or absence of a bare $\mu$ term depends on details of the ultraviolet completion so one might hope to address this issue there without making use of an explicit symmetry.  As for dimension 5 proton decay, the suppression by $\Lambda$ is not sufficient in itself but the proton lifetime depends on a number of factors \cite{Raby:2002wc} which can allow some room for adequate suppression.  One might also use family symmetries, as proposed in an F-theory context for instance in \cite{King:2010mq}, to do the job.  Both of these issues must be dealt with in a successful F-theory model for flipped $SU(5)$.

\section{Challenges for Realizing Flipped $SU(5)$ in $F$-Theory}
\label{sec:Ftheorychallenges}

We now turn to a discussion of flipped $SU(5)$ in the context of semi-local F-theory models.

\subsection{Engineering GUT-Higgs Fields}
\label{subsec:engineeringGUTHiggs}

Because $SU(5)\times U(1)_{\chi}$ naturally embeds into $SO(10)$, one way to engineer flipped $SU(5)$ models in F-theory is to realize an $SO(10)$ gauge group and explicitly break it to $SU(5)\times U(1)_{\chi}$ with internal flux.  The flux necessary to do this has the advantage that, unlike hypercharge flux, it does not split the gauge couplings at the high scale \cite{Jiang:2009za}.  There has been recent interest in building GUT models in this way and a number of semi-local and global constructions have been achieved \cite{King:2010mq,Chen:2010tp,Chen:2010ts,Chung:2010bn}.  Some of the constructions in \cite{Chen:2010ts} utilize internal fluxes not only to break $SO(10)\rightarrow SU(5)\times U(1)_{\chi}$ but also to further break this down to the MSSM.  As we are interested in alternatives to hypercharge flux in this note, we will insist in what follows on using GUT-Higgs fields and the flipped superpotential \eqref{flippedW} to break $SU(5)\times U(1)_{\chi}$ to the MSSM and lift the MSSM Higgs triplets.

One problem that has been noted by several authors \cite{Chen:2010tp,Chen:2010ts,Chung:2010bn} is that it is difficult to get the right spectrum including the GUT-Higgs fields.  While the MSSM matter multiplets organize nicely into $\mathbf{16}$'s of $SO(10)$ and the MSSM Higgs doublets and their triplet partners fit into a $\mathbf{10}$ of $SO(10)$, the GUT-Higgs $H$ and $\overline{H}$ do not fill out $SO(10)$ multiplets.  Rather, each must come from part of a $\mathbf{16}$ of $SO(10)$ and it is here that the problems arise.  In the presence of $N$ units of $U(1)_{\chi}$ flux, the net chirality of multiplets that descend from $\mathbf{16}$'s follows the pattern
\begin{equation}\begin{split}n_{\mathbf{\overline{5}}_{+3}} - n_{\mathbf{5}_{-3}} &= M+N \\
n_{\mathbf{10}_{-1}}-n_{\mathbf{\overline{10}}_{+1}} &= M \\
n_{\mathbf{1}_{-5}}-n_{\mathbf{1}_{+5}} &= M-N
\end{split}\end{equation}
where $M$ is the number of units of a suitable global $G$-flux that threads the matter curve.  Any excess of $\mathbf{10}_{-1}$'s or $\mathbf{\overline{10}}_{+1}$'s is accompanied by an excess of $\mathbf{\overline{5}}_{+3}$/$\mathbf{5}_{-3}$'s or $\mathbf{1}_{-5}/\mathbf{1}_{+5}$'s.  

To avoid introducing extra exotics, then, it becomes necessary to assume that $H$ and $\overline{H}$ simply arise as a vectorlike pair on a single matter curve.  This has two consequences.  First, any $U(1)$ symmetry from Table \ref{chargetable} that happens to be preserved must give opposite charge to $H$ and $\overline{H}$, meaning that  $p+q=0$ and the global $U(1)$ charges are simply proportional to $U(1)_{\chi}$.  Second, we must address why the GUT-Higgs fields are light or, if they sit at the KK scale, how such massive fields could possibly acquire nonzero vevs.

To make $H$ and $\overline{H}$ light, one could start by requiring the matter curve on which they live to support a vector-like pair of the appropriate zero modes.  Even then, one could not be certain that this pair does not become massive by coupling to moduli fields that acquire large nonzero vevs.  Alternatively, one could imagine starting with $H$ and $\overline{H}$ as two modes among the KK tower of $\mathbf{10}_{-1}$ fields and effectively bringing down their mass through an $SO(10)$ singlet $\Phi$ and a superpotential of the form
\begin{equation}W \supset \lambda_{16}\Phi\times \mathbf{16}_H\times \mathbf{\overline{16}}_H + M_{KK}\mathbf{16}_H\times\mathbf{\overline{16}}_{\overline{H}}\end{equation}

In general, the masses of different components of the $\mathbf{16}_H/\mathbf{\overline{16}}_{\overline{H}}$ will differ by order one multiples of $M_{KK}$.  A suitable vev of $\Phi$ could therefore render the $\mathbf{10}_H/\mathbf{\overline{10}}_{\overline{H}}$ pair very light while leaving the remaining components near the KK scale.  Of course, $\Phi$ will in general couple to all KK modes on the $\mathbf{16}$ matter curve and there is no reason for this cancellation to occur only in the $\mathbf{10}_H\mathbf{\overline{10}}_{\overline{H}}$ direction and not in the others. If fact, such a cancellation is not well motivated and will likely lead to an additional enormous fine-tuning in the theory that flipped $SU(5)$ was engineered to avoid.  Proceeding in this way seems quite cumbersome and will require many new assumptions.

An alternative to this, which seems particularly attractive, is to engineer $SU(5)\times U(1)_{\chi}$ directly.  In particular, one realizes an $SU(5)$ gauge group with a $U(1)_{\chi}$ following the construction of semi-local $SU(5)$ GUT models \cite{Marsano:2009gv,Blumenhagen:2009yv,Marsano:2009wr,Grimm:2009yu} and attempts to construct the global completion in such a way that the $U(1)_{\chi}$ survives as an honest gauge symmetry.  The nature of $U(1)$'s away from the GUT-divisor is rather subtle but there has been substantial recent progress \cite{Grimm:2010ez,Marsano:2010ix} towards understanding them.  The advantage of this approach is that one can engineer $H$ and $\overline{H}$ directly on separate $\mathbf{10}$ curves.  An example of a simple model that achieves this is constructed in Appendix B.

Unfortunately, two things remain to be resolved before realistic models can be built in this way.  First, one must be wary that global fluxes may lift $U(1)_{\chi}$ in the same way that hypercharge flux lifts $U(1)_Y$.  A necessary condition for this will be that  $U(1)_{\chi}$ be non-anomalous, which leads to the second issue.  While there has been progress towards understanding global fluxes in $F$-theory models \cite{Marsano:2010ix}, there is no simple procedure at the moment for counting the number of ($U(1)_{\chi}$-charged) $SU(5)$ singlet fields in a given model.  For flipped $SU(5)$, it is crucial that the number of such fields is 3 so this must be addressed before further progress can be made in this direction.

\subsection{Symmetries}

Next, we must be sure to incorporate enough symmetry to address the phenomenological problems discussed earlier in this note.  For dimension 4 $R$-parity violation, discrete symmetries seem unavoidable.  Engineering these can be technically challenging and the only serious attempt we are aware of in any context is in \cite{Hayashi:2009bt}.  That example already displays several pitfalls as even getting a reasonable number of generations seems difficult.  This seems like a technical hurdle, though, with no conceptual obstruction blocking the way.

More troublesome is the $\mu$ problem which, as we have seen, requires a $U(1)_R$ symmetry to resolve in a satisfactory way.  Unfortunately, semi-local $F$-theory models do not afford us this luxury.  As reviewed in Appendix A, these models descend from an 8-dimensional $E_8$ gauge theory with ${\cal{N}}=1$ supersymmetry in the presence of a background field configuration that breaks $E_8\rightarrow SU(5)$.  The 8-dimensional theory possesses a $U(1)_R$ symmetry and, further, additional $R$-symmetries could in principle follow from internal isometries of the compactification manifold that takes us from 8 down to 4 dimensions.  Because we retain only ${\cal{N}}=1$ supersymmetry in 4-dimensions, though, the supercharges are scalars with respect to the (twisted) internal isometries so only the remnant of the 8-dimensional $U(1)_R$ remains as a candidate.  This symmetry, however, is broken explicitly by the background field configuration so that no continuous $R$-symmetry remains to control physics at the KK scale{\footnote{Strictly speaking, there is a combination of topological and $R$-symmetries that remain unbroken by the scalar vev of the Higgs bundle.  This is broken by the flux part of the Higgs bundle.  Further, the 4-dimensional fields do not carry definite charge under this symmetry, so it could not constrain their physics anyway.}}.

We view this $\mu$ problem as the most glaring issue for engineering flipped $SU(5)$ models in $F$-theory.  It may be possible to avoid it phenomenologically with some intricate model building.  Finding a scenario that can be easily realized within the rigid framework of $F$-theory, though, will be challenging.

\subsection{Summary}

We started by looking to flipped $SU(5)$ as a means to avoid some problems with minimal $SU(5)$ models in $F$-theory but flipped $SU(5)$ has a number of issues as well.  Whether the situation is better or worse depends on one's taste but, in our opinion, the advantages of flipped $SU(5)$ are outweighed by the weaknesses.  We stress, however, that all of the issues discussed here rely on the explicit use of GUT-Higgs fields to break $SU(5)\times U(1)_{\chi}$ and lift the Higgs triplets.  Models based on $SO(10)$ with all GUT-breaking via internal fluxes \cite{Chen:2010ts} do not suffer from any of the problems related to GUT-Higgs fields, including their origin and their knack for generating large contributions to the $\mu$ term.

\section{Concluding Remarks}

In this work we explored the possibility of engineering a flipped $SU(5)$ model in F-theory. In particular, we show that a significant $\mu$ parameter ($\gtrsim 10^{10}$ GeV) is unavoidable in any flipped $SU(5)$ model without an $R$-symmetry. Since no four-dimensional $R$-symmetries control the superpotential in F-theory GUTs, we conclude that Flipped $SU(5)$ is not a viable mechanism to break the GUT group and solve doublet-triplet splitting in F-theory.

We also explored other problems, although not as deadly as the lack of an $R$-symmetry, that can arise when trying to embed Flipped $SU(5)$ in a UV completion that has a conserved $R$-symmetry. In the process, it is determined that at least one discrete symmetry is phenomenologically required to prevent severe $R$-parity violation, and that this symmetry cannot descend from a continuous $U(1)$ symmetry -- which is an issue when realizing discrete symmetries in some string constructions. Additionally, if one wishes to explain the scale of the neutrino masses this will necessarily re-introduce a $\mu$-problem regardless of whether or not there is an $R$-symmetry. However, this $\mu$ -problem can be solved if GUT sized $SU(5)$ singlet vevs are included in the theory.

Finally, we described some challenges specific to building F-theory models of flipped $SU(5)$.  Assuming that a suitable solution to the $\mu$ problem described above can be found, we mentioned some issues associated with engineering GUT-Higgs fields that have been noted before \cite{Chen:2010tp,Chen:2010ts,Chung:2010bn} when one tries to obtain flipped $SU(5)$ from $SO(10)$ models in F-theory.  A direct engineering of $SU(5)\times U(1)_{\chi}$ is one possible alternative to this and we constructed a sample semi-local model of this sort in Appendix \ref{sec:compact}.  Several global issues must be addressed related to $U(1)_{\chi}$ and $SU(5)$ singlets, though, in order for a model of this type to be truly successful.  The problem of engineering flipped $SU(5)$ in F-theory thus faces a variety of challenges and will require careful model building and new technical advances in our understanding of global F-theory vacua to overcome.

\section*{Acknowledgements}

We are grateful to T.~Cohen, M.~Cvetic, G.~Kane, S.~King, P.~Langacker, G.~Leontaris, A.~Pierce, and  S.~Sch\"afer-Nameki  for stimulating discussions.  EK would like to thank the Enrico Fermi Institute for hospitality during the course of this work while JM is grateful to the Michigan Center for Theoretical Physics and the organizers of the KITP workshop "Strings at the LHC and in the Early Universe".  Both of us are grateful to the KITP and organizers of the SVP Spring Meeting for hospitality.  The work of EK is supported by DOE Grant DE-FG02-95ER-40899 and a String Vacuum Project Graduate Fellowship, funded through NSF grant PHY/0917807. The work of JM is supported by DOE grant DE-FG02-90ER-40560.

\appendix

\appendix

\section{Semi-local F-theory Models}
\label{sec:Ftheory}

In this Appendix, we would like to address the presence or absence of (non-accidental) $R$-symmetries in semi-local F-theory models.  For this, recall that F-theory describes nonperturbative configurations of intersecting 7-branes in type IIB string theory.  Non-Abelian gauge theories can be engineered when several branes coincide.  To describe the gauge degrees of freedom, it is sufficient for many purposes to consider the worldvolume theory on the branes, which is sensitive to some aspects of the local geometry but is largely independent of global details of the compactification.  In all known examples for engineering SUSY GUTs, the brane worldvolume theory can be described as the maximally supersymmetric $E_8$ Yang-Mills theory in 8-dimensions compactified down to 4-dimensions in the presence of a nontrivial configuration for the internal gauge field and an adjoint scalar field.  Aspects of the local geometry manifest themselves by specifying this configuration, which breaks $E_8$ down to the GUT group while giving spatially varying masses to internal wave functions that localize bifundamental matter to "matter curves".  When we refer to a semi-local F-theory model, we mean precisely this 8-dimensional $E_8$ gauge theory with accompanying internal field configuration, which is often referred to as a Higgs bundle{\footnote{It should be noted that the assumption of a global $E_8$ as a starting point may not be general enough to capture all possible F-theory realizations of supersymmetric GUT models.  To date, however, we know of no examples of F-theory compactifications, or even local models that manage to describe the geometry along the entire GUT divisor (as opposed to just a single coordinate patch), that engineer a GUT while avoiding this global $E_8$ structure.}}.

In general, $R$-symmetries of models obtained by compactifying brane worldvolumes descend either from $R$-symmetries of the original brane theory or internal symmetries of the compactification.  This makes it easy to see that there are no continuous $R$-symmetries present in semi-local F-theory models; the theory undergoes a twisting that removes any $R$-symmetries that could have descended from the compactification while the Higgs bundle explicitly breaks the $U(1)_R$ of the original 8-dimensional theory.  In the following, we describe the twisting and the $R$-symmetry of the underlying 8-dimensional theory in a bit more detail to make this point clear to readers not familiar with the structure of F-theory models.  This discussion very closely follows that of \cite{Beasley:2008dc} with only a few minor emphases on $R$-symmetries added.  For a more detailed discussion of the worldvolume theory, including not just the twisting but also an explicit construction of the action, the interested reader is referred to \cite{Beasley:2008dc}.

\subsection{Brane Worldvolume Theory}

The worldvolume theory on a stack of 7-branes is a dimensional reduction of the 10-dimensional maximally supersymmetric Yang-Mills theory, whose field content consists of a 10-dimensional vector $A_I$ ($I=0,\ldots,9)$ and an $SO(9,1)$ Majorana-Weyl spinor of positive chirality ($\mathbf{16}_+$), $\Psi_A$.  The supercharges of this theory organize themselves into the same representation, $\mathbf{16}_+$, as the fermions.  In 8-dimensions, we obtain an 8-dimensional vector $A_i$ ($i=0,\ldots,7$), a complex scalar $\Phi=A_8+iA_9$, and an $SO(7,1)$ chiral spinor $S_+$ (along with its anti-chiral conjugate $S_-$).
The $R$-symmetry of the 8-dimensional theory is the $U(1)$ that descends from $SO(9,1)$ under the reduction
\begin{equation}SO(9,1)\rightarrow SO(7,1)\times U(1)_R\end{equation}

In F-theory applications, this 8-dimensional theory is compactified on a complex surface $S$, leaving us with a field theory 4-dimensions.  Because $S$ has a nontrivial canonical bundle in general, objects that transform as spinors under local $SO(4)$ rotations are not globally well-defined; rather, they are transformed by nontrivial transition functions as one moves from coordinate patch to coordinate patch.  The lack of a globally well-defined spinor, which is needed to define 4-dimensional supercharges, clashes with our knowledge that the F-theory compactifications under study manifestly preserve ${\cal{N}}=1$ supersymmetry in 4-dimensions.  This tension tells us that the 7-brane worldvolume theory is necessarily twisted, meaning that its coupling to the background metric is altered in a way that effectively replaces the local $SO(4)$ rotation group with a combination of $SO(4)$ and $U(1)_R$.  In fact, as described in \cite{Beasley:2008dc}, the twisting should respect the K\"ahler structure of $S$, which is only preserved by a $U(2)$ subgroup of $SO(4)$.  This means that the twisting can be specified by a particular embedding of $U(1)_R$ into $U(2)\subset SO(4)$.
To see the effect the twisting, consider first the way that 8-dimensional spinors of the theory organize into representations of $SO(3,1)\times U(2)\times U(1)_R=[SU(2)\times SU(2)]\times U(2)\times U(1)_R$.  Specifying a $U(2)$ representation by an $SU(2)$ representation and $U(1)$ charge, one has that under the decomposition
\begin{equation}SO(7,1)\times U(1)_R\rightarrow SO(3,1)\times U(2)\times U(1)_R\end{equation}
the 8-dimensional chiral spinor $(S_+,+1/2)$ reduces as
\begin{equation}\left(S_+,+\frac{1}{2}\right)\rightarrow \left[(\mathbf{2},\mathbf{1}),\mathbf{2}_0,+\frac{1}{2}\right]\oplus \left[(\mathbf{1},\mathbf{2}),1_{+1}\oplus 1_{-1},+\frac{1}{2}\right]\end{equation}
In order to obtain one 4-dimensional chiral supercharge that transforms as a scalar under the modified internal rotation group, one must replace the generator $J$ of the $U(1)\subset U(2)$ with one of the combinations
\begin{equation}J_{\text{top}}=J\pm 2R\end{equation}
where $R$ is the generator of $U(1)_R$.  Both of these lead to equivalent theories.  Taking the + sign, the $SO(3,1)\times U(2)\times U(1)_R$ transformation properties of $S_+$ become
\begin{equation}\left(S_+,+\frac{1}{2}\right) \rightarrow \left[(\mathbf{2},\mathbf{1}),\mathbf{2}_{+1},+\frac{1}{2}\right]\oplus \left[(\mathbf{1},\mathbf{2}),1_{+2}+1_0,+\frac{1}{2}\right]\end{equation}
where now the subscript refers to the $J_{top}$ charge.  The $[(\mathbf{1},\mathbf{2}),1_0,+1/2]$ component gives rise to an anti-chiral supercharge in 4-dimensions that is globally well-defined on $S$.  Decomposing the supercharges of the 8-dimensional theory in this way, these scalars give the supercharges of the resulting ${\cal{N}}=1$ theory.  Because the supercharges are scalars under the "twisted" internal rotation group, no $R$-symmetry can arise from there.  The $U(1)_R$ that descends from the $R$-symmetry of the 8-dimensional theory, however, remains a global symmetry.  This is the origin of a $U(1)_R$ symmetry in the 4-dimensional theory with respect to which the chiral supercharges carry charge $-\frac{1}{2}$.

Turning to the matter fields, the normalization of $U(1)\subset U(2)$ is such that it acts as $-p$ on holomorphic $p$-forms and $p$ on anti-holomorphic $p$-forms \cite{Beasley:2008dc}.  This means that $J_{top}$ is really a sort of topological charge, even though the theory itself is not topological.  Following \cite{Beasley:2008dc}, we write the fields that descend from 8-dimensional chiral fermions in the following way, where we specify again the $SO(3,1)\times U(2)\times U(1)_R$ representations for clarity (here $m/\bar{m}$ denote holomoprhic/antiholomorphic form indices)
\begin{equation}\Psi_A\rightarrow \left\{\begin{array}{cc}\psi^{\alpha}_{\bar{m}} &\sim \left[(\mathbf{2},\mathbf{1}),\mathbf{2}_{+1},+\frac{1}{2}\right] \\
\bar{\chi}^{\dot{\alpha}}_{\bar{m}\bar{n}} &\sim \left[(\mathbf{1},\mathbf{2}),\mathbf{1}_{+2},+\frac{1}{2}\right] \\
\bar{\eta}^{\dot{\alpha}} &\sim \left[(\mathbf{1},\mathbf{2}),\mathbf{1}_0,+\frac{1}{2}\right]\end{array}\right.\label{twistedfermions}\end{equation}
along with their conjugates.  

So far we have only considered fermion fields.  The 8-dimensional scalar $\Phi$, begins life as an $SO(7,1)$ singlet that carries $U(1)_R$ charge $+1$.  After the twisting, its $SO(3,1)\times U(2)\times U(1)_R$ representation is
\begin{equation}\Phi\sim \left[(\mathbf{1},\mathbf{1}),\mathbf{1}_{+2},+1\right]\label{twistedbosons1}\end{equation}
We also get scalars $A_m/A_{\bar{m}}$ with holomorphic/antiholomorphic indices $m/\bar{m}$ from dimensional reduction of the 8-dimensional vector.  The scalar $A_{\bar{m}}$ has $SO(3,1)\times U(2)\times U(1)_R$ representation
\begin{equation}A_{\bar{m}}\sim \left[(\mathbf{1},\mathbf{1}),\mathbf{2}_{+1},0\right]\label{twistedbosons2}\end{equation}

The action of the twisted 8-dimensional gauge theory and its dimensional reduction are studied in detail in \cite{Beasley:2008dc}.  There, it is noted that the fermions \eqref{twistedfermions} and bosons \eqref{twistedbosons1} \eqref{twistedbosons2} naturally pair up into ${\cal{N}}=1$ chiral multiplets $(A_{\bar{m}},\psi^{\alpha}_{\bar{m}})$, $(\phi_{mn},\chi^{\alpha}_{mn})$, and a vector multiplet $(A_{\mu},\eta_{\alpha})$.  In this language, the 4-dimensional superpotential can be written as
\begin{equation}W=\int_S\,d^2\theta\,\text{tr}\left(F^{(0,2)}\wedge \phi\right)\label{appsup}\end{equation}
where we denote chiral superfields by their lowest components and $F_S^{(0,2)}=\bar{\partial}_AA+A\wedge A$ is the $(0,2)$ field strength on $S$.  We note that, by virtue of the integral over $S$ being topological, it is necessarily invariant under the topological charge $J_{top}$.  It is also easy to see that it is invariant under $U(1)_R$.  After all, $F^{(0,2)}$ is $R$-invariant, $\phi$ carries $R$-charge +1, and, as we have seen, each 4-dimensional supercoordinate carries $R$-charge $-\frac{1}{2}$ in the present normalization.  Invariance under $U(1)_R$ is not a surprise; it is a consequence of the fact that the 4-dimensional theory inherits the $U(1)_R$ symmetry of the 8-dimensional theory that we started with.

\subsection{Higgs Bundle}

This is not the full story, though.  To construct a semi-local GUT model we must add to this 8-dimensional theory a nontrivial configuration for both the scalar field $\phi$ and the internal field strength $F_S$.  This configuration must satisfy the BPS equations{\footnote{Throughout most of the literature, only Abelian configurations in which $[\phi,\bar{\phi}]=0$ are considered.  There, the flux $F_S$ must satisfy $\omega\wedge F_S=0$.}}
\begin{equation}F_S^{(0,2)}=F_S^{(2,0)}\qquad \partial_A\phi = \partial_A\bar{\phi}=0\qquad \omega\wedge F_S + \frac{i}{2}[\phi,\bar{\phi}]=0\label{higgsbps}\end{equation}
The field $\phi$ carries nonzero $U(1)_R$ and $U(1)_{top}$ charges so only a single linear combination survives.  This is the combination for which the superfield $\phi$ in \eqref{appsup} carries charge 0 while the superfield $A_{\overline{m}}$ and covariant derivatives $\partial_A$ carry charge 1.

The Higgs bundles of interest always have nonzero $F_S^{(2,0)}$, which carries charge 2 under this symmetry, so one might conclude that even this symmetry is broken.  We want to be a little careful about this because the dependence of Yukawa couplings on fluxes it is not always clear.  For instance, it is known that while the spectrum depends on the gauge flux, Yukawa couplings essentially do not \cite{Cecotti:2009zf}.  On the other hand, $F_S^{(2,0)}$ arises in F-theory from $G$-flux, which may or may not impact the Yukawas.  One might argue that we can only be sure of which symmetries control the superpotential in the limit of vanishing $F_S^{(2,0)}${\footnote{This seems kind of nonsensical because the spectrum jumps if we set $F_S^{(2,0)}$ to zero but our main point is that we are more comfortable with an argument that does not rely on explicit breaking of a $U(1)$ by $F_S^{(2,0)}$.}}.

Nevertheless, it is easy to see that the combination $U(1)_R$ and $U(1)_{top}$ that is preserved by $\phi$ cannot descend to a symmetry that constrains the 4-dimensional effective action for massless fields.  This is because of the coupled nature of the equations of motion for 4-dimensional fermions
\begin{equation}\begin{split}0&=\omega\wedge \partial_A\psi^{\alpha}+\frac{i}{2}[\bar{\phi},\chi^{\alpha}]\\
&=\omega\wedge \partial_A\psi^{\alpha}-\frac{i}{2}[\phi,\bar{\chi}^{\dot{\alpha}}] \\
&= \bar{\partial}\chi^{\alpha}-[\phi,\psi^{\alpha}]\\
&= \partial_A\bar{\chi}^{\dot{\alpha}}-[\bar{\phi},\bar{\psi}^{\dot{\alpha}}]
\end{split}\end{equation}
In the presence of a nontrivial expectation value for $\phi$, these equations imply position dependent masses that cause the internal wave functions to localize along "matter curves" where this expectation value vanishes{\footnote{As a meromorphic section on a complex surface, the vanishing locus of the expectation value for $\phi$ will generically consist of a collection curves.}}.  Because $\psi^{\alpha}$ and $\chi^{\alpha}$ are coupled by these equations, exciting a single mode on a matter curve corresponds to turning on nontrivial profiles for both of them.  The 4-dimensional field that results does not have a well-defined charge under, $U(1)_R$, $U(1)_{top}$, or the linear combination that is preserved by $\phi$, because $\psi^{\alpha}$ and $\chi^{\alpha}$ carry different charges under all of these symmetries.  For this reason, we do not expect any of these symmetries to control the superpotential for massless 4-dimensional fields, including both renormalizable operators and the nonrenormalizable ones that arise from integrating out KK modes.

\section{Engineering $SU(5)\times U(1)_{\chi}$ Directly}
\label{sec:compact}

As noted in section \ref{subsec:engineeringGUTHiggs}, constructing flipped $SU(5)$ models from $SO(10)$ GUTs in F-theory has some intrinsic difficulties, most notably realizing the GUT-Higgs fields without introducing new exotics into the spectrum.  For this reason, it may be preferable to engineer $SU(5)\times U(1)_{\chi}$ directly, without using $SO(10)$ as an intermediate structure.  Doing this gives up unification and introduces a fine-tuning associated with the closeness of $\alpha_1$ to $\alpha_2$ and $\alpha_3$ at the high scale.  Nevertheless, it is an alternative that may be interesting because, in such models, the $U(1)_{\chi}$ gauge boson will not be localized near the GUT branes but rather will correspond to a "bulk" closed string mode that can couple more readily to hidden sectors.  This may make such scenarios interesting for phenomenology.

In this Appendix, we present a sample semi-local construction of an $SU(5)\times U(1)_{\chi}$ model that also engineers a $U(1)_{PQ}$ symmetry{\footnote{By $U(1)_{PQ}$ symmetry we mean a $U(1)$ symmetry that allows the MSSM superpotential but forbids a bare $\mu$ term.}} capable of removing many, but not all, of the problematic nonrenormalizable operators involving the GUT-Higgs fields $H$ and $\overline{H}$.  This is the first explicit example we are aware of that realizes multiple $U(1)$ symmetries that generically contains no non-Kodaira type singularities{\footnote{Semi-local models with multiple $U(1)$'s were recently studied in \cite{Dudas:2010zb} but a further topologically tuning that isn't specified explicitly must be added in order to ensure the lack of non-Kodaira type singularities at isolated points where pairs of sections vanish.  The construction that we describe in the following is different from those and requires no additional tuning beyond the choice $\xi_2={\cal{O}}$.  Further, we explicitly build an example in which all objects that are used to construct the model are sections of bundles that admit holomorphic sections.}}
While we add fluxes to engineer a flipped $SU(5)$ spectrum, it should be straightforward to engineer an ordinary $SU(5)$ GUT as well in this setup.  For flipped $SU(5)$, it is necessary to engineer $SU(5)$ singlet fields as well as ensure $U(1)_{\chi}$ remains massless in the presence of flux.  Neither of these issues are sufficiently well-understood in global models to ensure that they can be solved but the parameter space of fluxes that we find in the semi-local model is large enough that it seems reasonable to expect that both of these shortcomings can be addressed in the future.

\subsection{Semi-Local Model}

We now turn to a semi-local model for an $SU(5)$ GUT that retains a $U(1)_{\chi}$ and $U(1)_{PQ}$ symmetry.  We provide only a brief review of semi-local models and how to construct them.  For a more complete discussion, see \cite{Marsano:2009gv}.

As described in Appendix \ref{sec:Ftheory}, the starting point is an $E_8$ gauge theory.  We must then introduce a Higgs bundle satisfying \eqref{higgsbps}.  This is done with a spectral cover ${\cal{C}}$ \cite{Donagi:2009ra}, which is a 5-sheeted cover of the complex surface, $S_{\rm GUT}$, on which the gauge theory is compactified.  To break $E_8\rightarrow SU(5)_{\rm GUT}$, the scalar $\phi$ must take values in the adjoint of the $SU(5)_{\perp}$ commutant of $SU(5)_{\rm GUT}$ inside $E_8$.  These can be parametrized by five eigenvalues that sum to zero
\begin{equation}\langle\phi\rangle\sim \begin{pmatrix}t_1 & 0 & 0 & 0 & 0\\0 & t_2 & 0 & 0 & 0\\ 0 & 0 & t_3 & 0 & 0\\ 0 & 0 & 0 & t_4 & 0 \\ 0 & 0 & 0 & 0 & t_5\end{pmatrix}\qquad \sum_{i=1}^5 t_i=0\end{equation}
Roughly speaking, one can think of each sheet of the cover as specifying one of the five eigenvalues $t_i$.  As one moves along $S_{\rm GUT}$, the $t_i$ are mixed under monodromy.  This is reflected in ${\cal{C}}$ by the manner in which the sheets are glued together.  In the absence of monodromy, the $U(1)^4$ Cartan of $SU(5)_{\perp}$ survives as a symmetry of the theory.  In the presence of monodromies, only those $U(1)$'s that are invariant survive.

Monodromies also affect the potential matter content of the theory.  All matter descends from the adjoint of $E_8$
\begin{equation}\mathbf{248}\rightarrow \mathbf{(24},\mathbf{1})\oplus (\mathbf{1},\mathbf{24})\oplus (\mathbf{10},\mathbf{5})\oplus (\mathbf{\overline{5}},\mathbf{10})\end{equation}
Without monodromy, we get 5 copies of the $\mathbf{10}$ that transform as a fundamental of $SU(5)_{\perp}$.  We use the $t_i$ to label these five copies, denoting them $\mathbf{10}_{t_i}$ for $i=1,\ldots,5$.  Similarly, we get 10 copies of the $\mathbf{\overline{5}}$ labeled as $\mathbf{\overline{5}}_{t_i+t_j}$ with $i\ne j$.  Finally, we get 24 singlets labeled as $\mathbf{1}_{t_i-t_j}$ with $i\ne j$.  One typically doesn't discuss singlets in the context of semi-local models because their wave functions do not localize on the GUT-branes.  $SU(5)$ singlets are therefore sensitive to global details of the geometry so it doesn't make sense to describe much about them in a semi-local setting other than their charges under any $U(1)$ factors that remain.

A generic monodromy group will mix all $t_i$'s.  This projects out all extra $U(1)$'s and leads to a spectrum with just one type of $\mathbf{10}$ and one type of $\mathbf{\overline{5}}$.  We want to realize extra $U(1)$ symmetries to we construct a Higgs bundle with a reduced monodromy group by using a factored spectral cover ${\cal{C}}$.  In order to realize both $U(1)_{\chi}$ and $U(1)_{PQ}$ and engineer both the MSSM superpotential and the flipped superpotential \eqref{flippedW}, there is in fact a unique factorization structure
\begin{equation}{\cal{C}}\rightarrow {\cal{C}}^{(a)}\times {\cal{C}}^{(d)}\times {\cal{C}}^{(e)}\end{equation}
where ${\cal{C}}^{(a)}$ has two sheets, ${\cal{C}}^{(d)}$ has two sheets, and ${\cal{C}}^{(e)}$ has one sheet.  The matter fields that one obtains and their charges under the two $U(1)$'s that survive are listed below
\begin{equation}\begin{array}{c|cc}\text{Field} & U(1)_{\chi} & U(1)_{PQ} \\ \hline
\mathbf{10}^{(a)}\equiv \mathbf{10}_{t_a} & 1 & 1 \\
\mathbf{10}^{(d)} \equiv \mathbf{10}_{t_d} & 1 & -1\\
\mathbf{10}^{(e)} \equiv \mathbf{10}_{t_e} & -4 & 0 \\
\mathbf{\overline{5}}^{(aa)}\equiv \mathbf{\overline{5}}_{t_{a_1}+t_{a_2}} & 2 & 2 \\
\mathbf{\overline{5}}^{(dd)}\equiv \mathbf{\overline{5}}_{t_{d_1}+t_{d_2}} & 2 & -2 \\
\mathbf{\overline{5}}^{(ad)}\equiv \mathbf{\overline{5}}_{t_a+t_d} & 2 & 0 \\
\mathbf{\overline{5}}^{(ae)}\equiv \mathbf{\overline{5}}_{t_a+t_e} & -3 & 1 \\
\mathbf{\overline{5}}^{(de)} \equiv \mathbf{\overline{5}}_{t_d+t_e} & -3 & -1
 \end{array}\end{equation}
Our identification of the first $U(1)$ as "$U(1)_{\chi}$" is natural once we identify the fields above with those of the MSSM in the following way
\begin{equation}\begin{split}
\mathbf{10}^{(a)}&\leftrightarrow F + H\\
\mathbf{\overline{10}}^{(d)} &\leftrightarrow \overline{H} \\
\mathbf{5}^{(aa)} &\leftrightarrow h \\
\mathbf{\overline{5}}^{(dd)} &\leftrightarrow \overline{h} \\
\mathbf{\overline{5}}^{(ae)} &\leftrightarrow \mathbf{\overline{f}}
\end{split}\end{equation}
To engineer the right spectrum, then, we need the following chiralities of zero modes on each matter curve
\begin{equation}\begin{array}{c|cccccccc} \text{Curve} & \mathbf{10}^{(a)} & \mathbf{10}^{(d)} & \mathbf{10}^{(e)} & \mathbf{\overline{5}}^{(aa)} & \mathbf{\overline{5}}^{(dd)} & \mathbf{\overline{5}}^{(ad)} & \mathbf{\overline{5}}^{(ae)} & \mathbf{\overline{5}}^{(de)} \\ \hline 
\text{Chirality} & 4 & -1 & 0 & -1 & 1 & 0 & 3 & 0
\end{array}\label{desiredspectrum}\end{equation}

\subsubsection{Spectral Cover}

To construct such a model explicitly, we need a factored spectral cover.  The spectral cover lives in an auxiliary space that is the total space of the canonical bundle over $S_{\rm GUT}$.  We refer the reader to \cite{Marsano:2009gv} for details about this and only summarize the construction here.  We write a factored cover as
\begin{equation}{\cal{C}} = {\cal{C}}^{(a)}{\cal{C}}^{(d)}{\cal{C}}^{(e)}\end{equation}
with
\begin{equation}\begin{split}{\cal{C}}^{(a)} &= a_2V^2+a_1UV+a_0U^2 \\
{\cal{C}}^{(d)} &= d_2V^2+d_1UV+d_0U^2 \\
{\cal{C}}^{(e)} &= e_1V+e_0U\end{split}\end{equation}
Here, the $a_m$, $d_n$, and $e_p$ are sections of the bundles
\begin{equation}\begin{array}{c|c}\text{Section} & \text{Bundle} \\ \hline
a_m & \eta-(m+3)c_1-\xi_1-\xi_2 \\
d_n &= \xi_1+(2-m)c_1 \\
e_p &= \xi_2 + (1-p)c_1
\end{array}\end{equation}
where $c_1$ is short for the anti-canonical bundle of $S_{\rm GUT}$, $K_{S_{\rm GUT}}^{-1}$.  We can choose the bundle $\eta$, which encodes the manner in which $S_{\rm GUT}$ is embedded into a global model, as well as the bundles $\xi_1$ and $\xi_2$.  The traceless condition on ${\cal{C}}$, which amounts to ensuring that it specifies an $SU(5)$ bundle rather than a $U(5)$ one, becomes
\begin{equation}e_0d_0a_1+e_0a_0d_1+d_0a_0e_1=0\end{equation}
We choose to solve this in a very particular way that is tailored for our ultimate choice of the complex surface $S_{\rm GUT}$.  First, we take $\xi_2$ to be a trivial bundle so that we can set $e_1=$.  Because of this, we hereafter refer to $\xi_1$ simply as $\xi$
\begin{equation}\xi\equiv \xi_1,\quad \xi_2={\cal{O}},\quad e_1=1\end{equation}
Now, we define new sections $A,B,C$ and set
\begin{equation}\begin{split}a_0 &= d_1B^2C-e_0a_1\\
d_0 &= a_1A^2C-e_0d_1 \\
e_0 &= ABC
\end{split}\end{equation}
With this parametrization, we are free to choose a bundle $\chi$ for the section $C$.  The spectral cover now takes the form
\begin{equation}{\cal{C}} = b_5V^5 + b_4V^4U+b_3V^3U^2 + b_2V^2U^3+b_1VU^4+b_0U^5\end{equation}
where
\begin{equation}\begin{split}
b_5 &= a_2d_2\\
b_4 &= a_1d_2+a_2(d_1+d_2ABC)\\
b_3 &= a_1(d_1+a_2A^2C)+d_1d_2B^2C\\
b_2 &= C\left[ d_1^2B^2+A(a_1+a_2ABC)(a_1A-d_1B)+d_2AB^2C(d_1B-a_1A)\right]\\
b_1 &= 0 \\
b_0 &=-A^2B^2C^3(a_1A-d_1B)^2
\label{thebs}
\end{split}\end{equation}
We will tailor our construction so that it can be embedded into Calabi-Yau 4-folds based on the geometries of \cite{Marsano:2009ym}{\footnote{These geometries were constructed to satisfy a topological condition \cite{Buican:2006sn,Beasley:2008kw,Donagi:2008kj} that allows GUT-breaking via $U(1)_Y$ flux.  While we will not utilize this method of GUT-breaking, we still use the geometries of \cite{Marsano:2009ym} because of their relative simplicity.}}.  There, $S_{\rm GUT}$ is a $dP_2$ surface, whose second homology is generated by a hyperplane class, $h$, and two exceptional curves, $e_1$ and $e_2${\footnote{The nonzero intersections are $h^2=1$ and $e_i^2=-1$.  All other intersections vanish.  We hope that context will avoid any confusion between the hyperplane class, $h$, and the up-type Higgs multiplet, which we also refer to as $h$.}}.  In terms of these, $c_1$ is simply
\begin{equation}c_1 = 3h-e_1-e_2\label{c1choice}\end{equation}
while, in the geometries of \cite{Marsano:2009ym}, $\eta$ is given by
\begin{equation}\eta = 17h-6(e_1+e_2)\label{etachoice}\end{equation}
Finally, we must be careful about our choices of $\xi$ and $\chi$ in order to ensure that the bundles associated to all sections really do admit holomorphic sections.  To that end, we take
\begin{equation}\xi=h-e_1\qquad \chi=h\label{xichichoices}\end{equation}
To see that this is ok, we now list all sections that appear in \eqref{thebs}, the general bundles of which they are sections, and the specific bundles for the choices \eqref{c1choice}, \eqref{etachoice}, and \eqref{xichichoices}
\begin{equation}\begin{array}{c|cc}\text{Section} & \text{General Bundle} & \text{Bundle in our }dP_2\text{ Construction} \\ \hline
a_2 & \eta-5c_1-\xi & h-e_2\\
a_1 & \eta-4c_1-\xi &4h-e_1-2e_2 \\
a_0 & \eta-3c_1-\xi & 7h-2e_1-3e_2\\
d_2 & \xi & h-e_1\\
d_1 & c_1+\xi & 4h-2e_1-e_2\\
d_0 & 2c_1+\xi & 7h-3e_1-2e_2\\
e_1 & {\cal{O}} & {\cal{O}} \\
e_0 & c_1 & 3h-e_1-e_2 \\
A & -\frac{1}{2}(\eta+\chi)+3c_1+\xi & h-e_2 \\
B & \frac{1}{2}(\eta-\chi)-\xi-2c_1 & h-e_1 \\
C & \chi & h 
\end{array}\end{equation}
For our specific choice, all of the bundles admit holomorphic sections.  We now list the classes of all matter curves inside $S_{\rm GUT}$
\begin{equation}\begin{array}{c|c|c|c|c}
\text{Field} & \text{Origin} & \text{Equation for Matter Curve in }dP_2 & \text{Homology Class} & \text{Class for our choices} \\ \hline
\mathbf{10}^{(a)} & {\cal{C}}^{(a)} & a_2 & \eta-5c_1-\xi & h-e_2 \\
\mathbf{10}^{(d)} & {\cal{C}}^{(d)} & d_2 & \xi & h-e_1 \\
\mathbf{10}^{(e)} & {\cal{C}}^{(e)} & \ast & \ast & \ast \\
\mathbf{\overline{5}}^{(aa)} & {\cal{C}}^{(a)}-{\cal{C}}^{(a)} & a_1 & \eta-4c_1-\xi & 4h-e_1-2e_2\\
\mathbf{\overline{5}}^{(dd)} & {\cal{C}}^{(d)}-{\cal{C}}^{(d)} & d_1 & c_1+\xi & 4h-2e_1-e_2 \\
\mathbf{\overline{5}}^{(ad)} & {\cal{C}}^{(a)}-{\cal{C}}^{(d)} & (a_2d_1+a_1d_2)+C(a_2A+d_2B)^2& \eta-4c_1 & 5h-2(e_1+e_2) \\
\mathbf{\overline{5}}^{(ae)} & {\cal{C}}^{(a)}-{\cal{C}}^{(e)} & d_1+a_2A^2C & c_1+\xi & 4h-2e_1-e_2 \\
\mathbf{\overline{5}}^{(de)} & {\cal{C}}^{(d)}-{\cal{C}}^{(e)} & a_1+d_2B^2C & \eta-4c_1-\xi & 4h-e_1-2e_2 \\
\end{array}\end{equation}

\subsubsection{Fluxes}

The next step is to introduce suitable fluxes to engineer the desired spectrum of zero modes \eqref{desiredspectrum}.  In a semi-local model this amounts to twisting the Higgs bundle as described in \cite{Donagi:2009ra}.  We will make use of several fluxes.  These include two non-universal fluxes that are only accommodated if we further specialize the spectral cover.  To that end, we set
\begin{equation}\begin{split}a_1 &= \alpha\tilde{\alpha}-d_2B^2C\\
d_1 &= \delta\tilde{\delta}-a_2A^2C\end{split}\end{equation}
We will abuse notation in what follows and use $\alpha,\delta$ to denote both the sections above and the bundles of which they are sections.  With this in mind, the fluxes that we introduce are
\begin{equation}\begin{split}\gamma_a &= n_a \left(2-p_a^*p_{a*}\right)\sigma\cdot {\cal{C}}^{(a)} \\
\gamma_d &= n_d \left(2-p_d^*p_{d*}\right)\sigma\cdot {\cal{C}}^{(d)} \\
\tilde{\Psi}_a &= \left\{ [V=e_0U]\cap \alpha\right\} - \alpha\cdot {\cal{C}}^{(e)} \\
\tilde{\Psi}_d &= \left\{[V=e_0U]\cap \delta\right\} - \delta\cdot {\cal{C}}^{(e)} \\
\tilde{\rho} &= \left[ {\cal{C}}^{(a)} - {\cal{C}}^{(d)}\right]\cdot\rho \\
\tilde{\mu} &= \left[ {\cal{C}}^{(a)} - 2{\cal{C}}^{(e)}\right]\cdot \mu \\
\tilde{\nu} &= \left[ {\cal{C}}^{(d)} - 2{\cal{C}}^{(e)}\right]\cdot \nu
\end{split}\label{thefluxes}\end{equation}
where $\rho$, $\mu$, and $\nu$ denote arbitrary classes in $H_2(dP_2,\mathbb{Z})$.  We also use $p_a$ to denote the projections $p_a:{\cal{C}}^{(a)}\rightarrow S_{\rm GUT}$ and similar for $p_d$.
All of these fluxes are constructed so that the net trace is zero but traces along individual components of ${\cal{C}}$ do not necessarily vanish.  The net flux that we construct must be supersymmetric, though.  The condition for supersymmetry that we impose is that the net flux, $\Gamma$, satisfies
\begin{equation}\omega\cdot p_{a*}\Gamma = \omega\cdot p_{d*}\Gamma = 0\end{equation}
for some $\omega$ in the K\"ahler cone of $S_{\rm GUT}=dP_2$.

To compute the spectrum from these fluxes, it is necessary to identify matter curves within the spectral cover ${\cal{C}}$ as described, for instance, in \cite{Marsano:2009gv}.  This is tedious but straightforward so we do not present the deteails here.  We simply note that, with the fluxes \eqref{thefluxes}, it is relatively easy to find a 4-parameter space of solutions that are supersymmetric and yield the proper spectrum \eqref{desiredspectrum}.  One sample solution from this space is
\begin{equation}\begin{split}n_u&=-1 \\
n_d &= 0 \\
\delta &= -h+7e_1-e_2 \\
\alpha &= -5h+3e_1+7e_2 \\
\rho &= -e_1 \\
\mu &= 0 \\
\nu &= 3h-5e_1
\end{split}\end{equation}
which satisfies the supersymmetry condition for $\omega = c_1 = 3h-e_1-e_2$.

\subsection{Comments on Global Embedding}

The semi-local model presented here is only a first step.  Embedding in a global model based on the geometries of \cite{Marsano:2009ym} is straightforward since we know how to lift sections of bundles on $dP_2$ to sections of bundles on the 3-fold described therein.  For sections that are not symmetric in $e_1$ and $e_2$ this can be a bit tricky but, as shown in \cite{Marsano:2009wr}, this can be dealt with.  One must worry about ensuring that $U(1)_{\chi}$ survives as an honest gauge symmetry, rather than being lost due to additional effects like those described in \cite{Hayashi:2010zp}, but from \cite{Grimm:2010ez,Marsano:2010ix} we know how to do this.  We need only lift the sections appearing in \eqref{thebs} to sections on the full 3-fold of \cite{Marsano:2009ym} and write a truncated Weierstrass from as in $y^2 = x^3 + {\cal{C}}$ \cite{Marsano:2010ix} with no additional terms.  That the fluxes \eqref{thefluxes} can be globally extended in this setting follows from the construction of \cite{Marsano:2010ix}.

What remains to be understood are two important ingredients.  The first is engineering the proper number of $SU(5)$ singlets, which is important because these become right-handed electrons in flipped $SU(5)$ models.  This will also ensure that $U(1)_{\chi}$ is non-anomalous, which is a necessary condition for having it remain as an honest massless gauge symmetry at low energies.  As we know for experience with hypercharge flux, though, this is not enough.  One must carefully ensure that the fluxes we use to induce chirality do not cause $U(1)_{\chi}$ to be lifted.  We hope that the parameter space of fluxes we have found is large enough to allow at least some choice that does not lift $U(1)_{\chi}$ but we have no way of saying for certain at the moment.  Further progress will require refined understanding of global fluxes and $U(1)$'s in F-theory beyond what is currently known.

\bibliographystyle{JHEP}
\renewcommand{\refname}{Bibliography}

\providecommand{\href}[2]{#2}\begingroup\raggedright\endgroup

\end{document}